# Inhomogeneous Superconductivity and the "Pseudogap" State of Novel Superconductors

**Vladimir Z. Kresin**

Lawrence Berkeley Laboratory, University of California at Berkeley,
CA 94720

**Yurii N. Ovchinnikov**

L.D. Landau Institute for Theoretical Physics, Russian Academy of
Sciences, 117334, Moscow, Russia

**Stuart A. Wolf**

Department of Materials Sciences, University of Virginia, Charlottesville, VA 22903

# Inhomogeneous Superconductivity and the "Pseudogap" State of Novel Superconductors

### Abstract

Novel superconducting compounds such as the high $T_c$ oxides are intrinsically inhomogeneous systems. An inhomogeneous structure is created by doping and the statistical nature of the distribution of dopants. Consequently, the critical temperature is spatially dependent: $T_c \equiv T_c(\mathbf{r})$.

Pair-breaking leads to a depression of $T_c$ and is a major factor leading to inhomogeneity . The "pseudogap" state is characterized by several energy scales : $T^*$, $T_c^*$, and $T_c \equiv T_c^{res.}$. The highest energy scale $(T^*)$ corresponds to the phase separation (at $T<T^*$) into mixed metallic-insulating structure. Especially interesting is the region $T_c^* > T > T_c$ : the compound contains superconducting "islands" embedded in the normal metallic matrix. As a result ,the system is characterized by normal conductance along with an energy gap structure, anomalous diamagnetism, unusual a.c. properties, isotope effect, and "giant" Josephson proximity effect. The energy gap may persist to temperature above $T_c^*$ caused by the presence of a charge density wave (CDW) or spin density wave (SDW) in the region $T>T_c^*$ by CDW, SDW, etc, whereas below $T_c^*$ the pairing also makes a contribution to the energy gap ($T_c^*$ is an "intrinsic" critical temperature). The values of $T^*$, $T_c^*$, $T_c \equiv T_c^{res.}$ depend on the compound and the doping level. The transition at $T_c$ into the dissipationless (R=0) macroscopically coherent state is of a percolation nature.

**Outline**

I. Introduction

Novel superconducting systems and, first of all, the high $T_C$ oxides, display many properties in the normal state above $T_C$, which are drastically different from those for conventional materials. This unusual normal (it is called "normal" because of finite resistance) state was dubbed the "pseudogap" state. The first observation of this state was reported in 1989, that is, shortly after the discovery of high $T_C$ superconductivity (Bednorz and Mueller, 1986). NMR measurements demonstrated the presence of an energy gap for spin excitations (Alloul et al. (1989), Warren et al. (1989)). Later, the presence of an energy gap was observed by the use of various techniques, such as tunneling, infrared, photoemission, heat capacity, etc.

One should stress that the presence of the gap is important, but, nevertheless, not the only special feature of the state above $T_C$. One can also observe peculiar magnetic transport and microwave properties as well as an isotope effect. An unusual "giant" Josephson proximity effect has also been observed. In the following review we will describe (Ch. II) the main properties of the "pseudogap" state.

The study of the "pseudogap" state has attracted a lot of attention. There are a number of experimental and theoretical papers describing interesting data and various theoretical models (see, e.g., reviews by Timusk and Statt (1999); Orenstein and Millis (2000); Tallon and Loram (2000)).

There is a general question about the definition of the "pseudogap" state. The consensus is that we are dealing with an unusual normal state, but such a definition is too vague. Some authors define the "pseudogap" state as a state above $T_C$ with an energy gap; this even resonates with the notation of the state. A more general definition implies that above $T_C$ the sample is in a peculiar state which is intermediate between fully superconducting and normal. Indeed, as in any normal metal, the sample displays a finite resistance. In addition, such a key feature as macroscopic phase coherence does not persists above $T_C$. However, one can observe some features typical for the superconducting state such as the energy gap, anomalous diamagnetism, and the isotope effect. It is interesting that the temperature scales (e.g., for the energy gap vs. diamagnetism) could be different. In addition, the manifestation of the "pseudogap" state depends on the doping level, being the strongest for the underdoped region.

Although the major focus of recent experimental studies has been on the high $T_C$ cuprates, the presence of the "pseudogap" state was reported also for other superconducting systems, such as borocarbides, bronzes, as well as for some more unusual conventional superconductors. We will discuss these properties below, Ch. VI.

During the last several years we have been greatly involved in the description of this peculiar state (Ovchinnikov et al., 1999, 2001, 2002; Kresin et al., 2000, 2003, 2004)

The structure of the paper is as follows. Ch. II contains a cross-cut description of the major experimental data. The concept of intrinsic inhomogeneity and the model describing the "pseudogap" state will be analyzed qualitatively in Ch. III. The different energy scales ($T^*$, $T_C^*$, $T_C$) will be introduced.

Ch. IV contains a detailed theoretical analysis and comparison with experimental data. The "giant" Josephson proximity effect is analyzed in Ch. V. Ch. VI focuses on other novel superconductors. And finally, Ch. VII contains concluding remarks and some open questions.

**II. "Pseudogap" state: main properties**

In general, novel superconducting systems and, in particular, cuprates are characterized by a finite resistance above $T_C^{res}$ and also macroscopic phase coherence disappear above $T_C^{res}$. However, a number of other features typical of the superconducting state, such as anomalous diamagnetism, energy gap, peculiar a.c. properties, etc. are observed above $T_C^{res}$. Of course, the presence of each isolated feature (e.g., energy gap, see below) might have several alternative explanations, but the theory should provide an unified explanation describing all relevant data. Note also, that the manifestations of various anomalous properties depends on the level of doping. The strongest anomalies occur in the underdoped region.

In this section we describe the main properties of the "pseudogap" state. As was mentioned above, the first experimental observation of this state was carried out with the use of the NMR technique. Later, many experimental methods have been employed; (some of them will be described below; see also reviews: Timusk and Statt, (1999)). Further below in this section (not in chronological order) we will follow some sequence which allows us to analyze the data in a consistent way.

## I.  Conventional superconductors: fundamentals.

In order to contrast the behavior of novel superconductors above $T_C^{res}$ with that for usual systems, we describe in this section several fundamental properties of conventional superconductors; see also Table I which directly contrasts the two classes.

The most fundamental feature is the anomalous diamagnetism (Meissner effect) observed below the critical temperature. As for the region above $T_C$, the magnetic response of conventional metals is relatively small and almost temperature independent.

The energy gap $\Delta$ is a fundamental microscopic parameter. As we know, $\Delta = 0$ in the normal phase and opens up at $T_C$. One should note, however, that the energy gap is indeed, an important parameter, but its presence is not a crucial factor for superconductivity. For example, one can observe "gapless superconductivity", (Abrikosov and Gor'kov (1961), caused by the pair-breaking effect, e.g., by the presence of localized magnetic moments.

As we know, superconductors have a finite resistance above $T_C^{res}$ whereas below the critical temperature they are in the dissipationless state (R = 0). In addition a.c. transport behaves in a peculiar way, namely, above $T_C$, with high accuracy (see, e.g., Landau and Lifshits,1960), ReZ = ImZ (Z is the surface impedance), contrary to this, below $T_C^{res}$ one can observe a strong inequality ReZ ≠ ImZ.

The superconducting state is also characterized by macroscopic phase coherence. For example, such a remarkable phenomenon as the Josephson effect is directly related to this feature.

Of course, because of fluctuations, the superconducting state can persists above $T_C$, but such a contribution can be properly analyzed. (see e.g., Larkin and Varlamov (2005)).

One can continue the list of properties contrasting the superconducting and normal state, but here we restrict ourselves to the aforementioned features (see also table I), because it is sufficient for our purpose.

## 2.2. Anomalous diamagnetism above $T_C^{res}$

Usual normal metals display relatively weak response to a small external magnetic field. Indeed, the electronic gas is characterized by small Pauli paramagnetism. The magnetic succeptibility of real metals consists of several contributions (see, e.g., Ashchroft (1976)) and the resulting response might be diamagnetic, but the total susceptibility is almost temperature independent.

The situation in the high $T_C$ cuprates appears to be drastically different. Unusual magnetic properties of the "pseudogap" state have been observed by several groups.

The scanning SQUID microscopy was used by Iguchi et al. (2001) to study the underdoped $La_{2-X}Sr_XCuO_4$ compound. This technique allows one to create a local magnetic image of the surface, its "magnetic" map. The critical temperature of the underdoped LSCO films was: $T_C^{res} \approx 18K$. A peculiar inhomogeneous picture has been observed (Fig.1): the film contains diamagnetic domains and their presence persists up to 80K (!). The total size of the diamagnetic regions is growing as the temperature is decreased (Fig. 2). The diamagnetic response appears to be strongly temperature dependent; this is a very unusual feature of the materials.

A strong temperature dependent diamagnetic response has been also observed by Bergemann et al. (1998) by using the torque magnetometry technique for the overdoped $Tl_2Ba_2CaO_{+\delta}$ compound above $T_C^{res} \approx 15K$. Like LSCO, the diamagnetic moment was also strongly temperature dependent (Fig. 3). Torque magnetometry was also employed recently by Wang et al. (2005) to study the Bi2212 compound. Similarly, diamagnetic response was observed. It is essential that the analysis ruled out fluctuations as a key source of the observed diamagnetism.

An interesting study of the YBCO compound has been carried out by Caretta et al. (2000) and by Lascialfari et al. (2002). The magnetization of the underdoped vs. optimally doped samples was measured by using special SQUID magnetometers. The observed diamagnetic response could be caused by superconducting fluctuations above $T_C$ (see e.g., Larkin and Varlamov (2005)). An anisotropic Ginzburg – Landau functional was employed to analyze the data. It has been concluded that fluctuations play an exceptional role and the diamagnetism above $T_C$ for the sample with optimum doping can be explained by their presence. However, and this is important for our analysis, their contributions are not sufficient to explain the data for the underdoped sample, and it is necessary to take into account the inhomogeneity of the structure (see below) to account for the observed diamagnetism. This effect is especially significant for the underdoped cuprates.

The strong diamagnetic response above $T_C^{res}$ has been reported not only for the cuprates, but also for other novel superconducting systems. We will describe these systems below (Ch. VI).

## 2.3. Energy gap

The presence of the energy gap above $T_C$ has been observed using various experimental methods. Even the title "pseudogap state" reflects the existence of the gap structure. In connection with this it is worth noting that this title is misleading, since we are dealing not with a "pseudogap" but with a real gap, that is, with real dip of the density states in the low energy region.

Let us start with tunneling spectroscopy which allows one to perform the most detailed and reliable study of the gap spectrum. The data obtained by Renner et al. (1998) for an underdoped Bi2212 crystal are shown in Fig. 4. The scanning tunneling microscopy (STM) of the crystals cleaved in vacuum was employed. One can see directly the dip in the density of states (energy gap) which persists above $T_C^{res} \approx 83K$ and stays up to $\approx 200K$ (!). One should note several interesting features of the data plotted in Fig. 4. First of all, the gap structure changes *continuously* from the superconducting region T<$T_C^{res}$ to the "pseudogap' state (T > $T_C^{res}$); there is not any noticeable change at $T_C^{res}$. This can be considered as an indication that the gap structure above $T_C^{res}$ is related to superconducting pairing.

The second interesting feature is that an increase in temperature affects the depth of the dip, but not the value of the gap, which is determined by its width. Such an unusual independence of the gap magnitude on temperature will be discussed in more detail below (Sec.IV.2 ). Tunneling measurements with similar results have been also carried out by Ekino et al. (1999).

The presence of an energy gap has been also determined using infrared spectroscopy. As a matter of fact, infrared spectroscopy was used in the pioneering works by Glover and Tinkham (1956), see Tinkham (1996); this was the first experimental observation of the gap in conventional superconductors.

The measurements of the c-axis conductivity in the underdoped YBCO compound and the corresponding analysis (Homes et al (1993) reveal the presence of a gap at temperatures above $T_C^{res}$ (Fig. 5 ). Again, it is interesting to note that the value of the gap ($\approx 400$cm$^{-1}$) is not noticeably affected by the transition $T_C^{res}$ to the dissipationless state at higher temperatures. The gap persists up to T $\approx$ 300K. It is Important that an increase in temperature leads to an decrease in depth of the gap, but the value of the gap (the width) remains almost constant and this is very similar to what is observed in tunneling (see above).

A special type of spectroscopy, the so-called "intrinsic tunneling spectroscopy" was developed by Suzuki et al. (1999) and by Krasnov et al. (2000). It is based on the fact that below $T_C^{res}$ the transport in the c-direction represents an intrinsic Josephson current (Kleiner et al., (1999), Scheekga et al. (1998)). This method can provide detailed information about the density of states. At the same time, intrinsic tunneling spectroscopy is sensitive to the stoichiometry of the sample, so that the method can be reliably used in the region near optimum $T_C$. According to the study of the temperature dependence of the I-V characteristics, one can distinguish two energy gaps, one is the pairing gap at $T < T_C$ and the second is a gap which differs from zero in a large temperature range below $T^*$ which is associated with the pseudogap.

The presence of an energy gap for spin excitations has been established using NMR. Actually, the "pseudogap" state was initially observed using this method. It has been observed in YBCO for different nuclei in both the Knight shift and spin-relaxation rate experiments: for $^{89}Y$ (Alloul et al. (1989), for $^{63}Cu$ (Warren et al. (1989), Walstedt et al. (1990), and for $^{17}O$ (Takigawa et al. (1989)).

Photoemission spectroscopy has also revealed the presence of an energy gap at $T > T_C \equiv T_C^{res}$ (see the reviews by Shen and Dessau (1995) and by Randeria and Campuzano (1997)). For example, it has been demonstrated that the energy gap persists in an underdoped sample of Bi2212 up to 95K; this temperature is much higher than $T_C^{res} = 79K$ (Loeser et al. (1997), see Fig. 6.

We previously described the data which presents direct spectroscopic observations of the gap structure above $T_C^{res}$. A gap in the spectrum can also be inferred from heat capacity data. One should note that measurements carried out by Loram et al. (1994) were one of the first observations of the "pseudogap" state. The measurements of the Sommerfeld constant $\gamma(T)$ display a loss of entropy caused by the gap structure. The data for the energy gap $\Delta(T)$ are derived from values of the electronic entropy $S(T, x)$ for $0.73 < x < 0.97$. Again, one can see that the energy gap persists for $T > T_C^{res}$, and the effect is especially strong for the underdoped samples. The substitution of Zn impurities for $Cu^{2+}$ on the $CuO_2$ planes leads to depression of $T_C^{res}$ and a strong impact on the value of $\gamma(T)$.

The $La_{2-X}Sr_XCuO_4$ compound was studied in detail by Momono et al. (1999). The data are similar to those of Loram et al. (1994). The authors also performed measurements for samples with Ni substitution.

**2.4. Transport properties.**

Let us now discuss other interesting features observed above the resistive $T_C^{res}$, in the "pseudogap" region.

As we know, the resistivity in the normal state at optimum doping is described by a linear temperature dependence $\rho_n \propto T$. However, a noticeable deviation from a simple linear law has been observed in the underdoped region. (Takagi et al. (1992), Hopfengartner et al. (1994), Sfar et al. (2005)).

The microwave properties and a.c. transport have been studied by Kusco et al, (2002). It was shown that above $T_C^{res}$, that is, in the normal state $ReZ \neq ImZ$, where Z is the surface impedance. This is an unusual property, since in ordinary normal metals with a high degree of precision, the real and imaginary part of the impedance are equal (see e.g., Landau and Lifshits, (1960)), that is $ReZ_n \cong ImZ_n$. The observed inequality $ReZ \neq ImZ$ is typically observed in superconducting materials.

In describing transport properties, one should also mention interesting data on the thermal Nernst effect (Xu et al.(2000),Wang et al.(2001),Ong and Wang (2004)). This effect is analogous to the Hall effect, but it is manifested by the appearance of an electric field created by an external magnetic field in the presence of a thermal gradient. Based on the data, the authors concluded that above $T_C^{res}$, in the "pseudogap" state, one can detect the presence of vortex-line excitations. Indeed, the value of the Nernst coefficient greatly exceeds that for any normal metal. On other hand, conventional superconductors of II type in the mixed state display a large value of this coefficient (Josephson (1965)). A careful analysis of the Nernst data (Wang et al.(2005)) reveals that the sharp rise in Nernst coefficient occurs very close to $T_c$ and dissappears about 10K above it in a region that can have large superconducting clusters which in principle can contain vortices.

As was noted above (Sec.2.2.), Wang et al. recently performed torque measurements which reveal diamagnetism above $T_c$; as a result, it was concluded, that the superconducting state persists above $T_c$.

One should also note that previous torque measurements (Bergemann et al, (1998)) on the overdoped Tl-based cuprate, also display diamagnetism above $T_c$( see Secs. 2.3. and 4.2) and the vortices were not detected even though this technique is quite sensitive to their presence.

The measurements of normal resistivity (Darmaoui and Jung (1998), Yan et al. (2000), Jung et al. (2000)) have revealed a strong inhomogeneity of YBCO and TBCCO samples.

The presence of two different phases and inhomogeneous structure of the order parameter has been demonstrated.

The so-called "Giant" Josephson proximity effect is another interesting phenomenon observed in the "pseudogap" region above $T_C$ (Bozovic et al.(2004). The films of $La_{0.85}Sr_{0.15}CuO_4$ ($T_C \approx 45K$) were used as electrodes whereas the underdoped LaCuO compound ($T_C \approx 25K$)formed the barrier which was prepared in the c-geometry (the coherence length $\xi_c \approx 4A$). the measurements were performed at $T_C' < T < 35K$, so that the barrier was in the "pseudogap" state. Since $T > T_C'$, we are dealing with the SNS junction. However, the Josephson current was observed for thicknesses of the barrier up to 200A(!). Such a "giant" effect can not be explained using conventional theory. We will discuss this effect in detail in
Ch. V.

**2.5. Isotope effect.**

Another interesting property of the "pseudogap" state is the strong isotope effect. This effect has been observed by Lanzara et al. (1999) for $La_{2-X}Sr_XCuO_4$ using x-ray absorption near-edge spectroscopy (XANES). The effect has been also observed by Temprano et al. (2000) for the $HoBa_2Cu_4O_8$ compound. The slightly underdoped $HoBa_2Cu_4O_8$ sample was studied by neutron spectroscopy. As we know (see, e.g., review by Mesot and Furrer (1997)), the opening of the gap which could be associated with the "pseudogap" affects the relaxation rate of crystal field excitations. The isotopic substitution $^{16}O \rightarrow ^{18}O$ leads to a drastic change in the value of the pseudogap temperature T$^*$ (T$^*\approx$ 170K → T$^*\approx$ 220K). Such a large isotope shift corresponds to a value of the isotope coefficient $\alpha^* = -2.2 \pm 0.6$. Note that, contrary to the typical superconductor, its value is negative.

### III. Inhomogenous superconductivity and the "pseudogap" phenomenon

As was described above, intensive experimental studies reveal a number of unusual features of the cuprates above the resistive transition T$_C^{res}$. We described anomalous diamagnetism which strongly depends on temperature, an energy gap structure, a strong inequality ReZ ≠ ImZ (Z is the surface impedance), a "giant" Josephson proximity effect, and an isotope effect on T$_C^*$.

All these phenomena can be explained in an unified way (Ovchinnikov et al., 1999; 2001; 2002; V. Kresin et al., 2003; 2004a; 2004b). We focus in this section on a qualitative picture and discuss some relevant experimental data. A more detailed theory will be described below (Sec. IV).

### 3.1. Qualitative picture

Consider an inhomogeneous superconductor, so that T$_C$=T$_C$(r). The system contains a set of superconducting regions "islands" embedded in a normal metallic matrix (Fig. 7). Properties of such system correspond to the "pseudogap" state. Indeed, the normal metallic matrix provides finite resistance whereas the existence of the supeconducting "islands" leads to an energy gap structure and the diamagnetic moment.

As was mentioned above (Sec. II), the presence of diamagnetic "islands" has been observed directly by Igushi et al. (2002). The superconducting "islands" are embedded in the normal metallic matrix. As a result, the proximity effect plays a crucial role. The proximity effect determines a minimum length scale of the superconducting regions which is of order of the coherence length $\xi_0$. Indeed, if a superconducting "island" has a size smaller than $\xi_0$, its superconducting state would be totally depressed by the proximity effect between the superconducting region and the normal metallic phase.

As temperature decreases towards $T_C \equiv T_C^{res}$, the size of the superconducting regions increases as does the number of "islands". The temperature $T_C \equiv T_C^{res}$ corresponds to the percolation transition, that is to the formation of a macroscopic superconducting region ("infinite cluster" in terms of the percolation theory see, e.g. Sklovskii and Efros (1984), Stuffer and Aharony (1992)), and to phase coherence and dissipationless superconducting phenomena. Note that
a similar inhomogeneous picture was later employed by
Mihailovic et al. (2002).

In conventional superconductors the resistive and Meissner transitions occur at the same temperature, $T_C$. The picture in the "pseudogap" state is different. The resistive and Meissner transition are split. The Meissner transition, (the appearance of the diamagnetism) occurs at $T_C^*$, whereas the resistive transition, that is, the transition to the macroscopic dissipationless state takes place at $T_C^{res}$, and $T_C^{res} < T_C^{Meis} \equiv T_C^*$.

It is important to note also that the "pseudogap" state in the region $T_C^{res} < T < T_C^*$ is not a phase coherent one; each superconducting "island" has its own phase. At $T \equiv T_C^{res}$ the macroscopic superconducting region is formed, and below $T_C^{res}$ we are dealing with macroscopically phase coherent phenomenon.

As was noted above, the presence of superconducting "islands" embedded in normal metallic matrix implies an inhomogeneity of the compound. There are two possible scenarios for such an inhomogeneous structure:

I.  inhomogeneous distribution of pair-breakers, and

II. inhomogeneous distribution of carriers leading to spatial dependence of the coupling constant.

Both scenarios lead to an inhomogeneous superconductivity but for the cuprates the first of them is dominant. Let us discuss it in more detail.

Pair breaking can be caused by localized magnetic moments (Abrikosov and Gor'kov, 1961). Qualitatively, the picture of pair-breaking can be visualized in the following way. A Cooper pair consists of two carriers with opposite spins (for singlet pairing; this is the case for both, s- or d- wave scenarios). A localized magnetic moment acts to align both spins in the same direction and this leads to pair-breaking.

It is well known that a pair-breaking effect leads to a depression in $T_C$. It is also known, that for d-wave pairing, non-magnetic impurities are also pair-breakers. Therefore, a non-uniform distribution of pair-breakers makes the critical temperature spatially dependent: $T_C \equiv T_C(\mathbf{r})$. Such a distribution is caused by the statistical nature of doping. The region which contains a larger number of pair-breakers is characterized by a smaller value of the local $T_C$.

A detailed theoretical analysis of the inhomogeneous structure caused by the presence of isolated non-magnetic and magnetic defects has revealed unusual properties for the density of states in superconductors (Vechter et al. (2003), Shyter et al. (2003); see also review by Balatsky et al. (2005)). They concluded that, strictly speaking, the density of states is finite everywhere in the superconducting gap. The density could be exponentially small, but different from zero even for an S-wave order parameter; the tail extends into the mean field gap. It was also shown (Zhu et al. (2005)) that isolated defects create long-range elastic deformations which lead to a local depression of the order parameter.

As was mentioned above, an inhomogeneity can be caused also by an inhomogeneous distribution of carriers (see Ovchinnikov et al. (2000)). However, since the inhomogeneous distribution of pair-breakers (dopants) appears to be the dominant source of inhomogeneity ,we focus below (see Sec. 4.3)on such a channel.

The percolative nature of the transition at $T = T_C^{res}$ is due to the statistical nature of doping. The picture is similar to that introduced in manganites which represents another family of doped oxides (see Gor'kov and Kresin (1998), Dzero et al. (2001); see also review by Gor'kov and Kresin (2004). Manganites (e.g., $La_{0.7}Sr_{0.3}MnO_3$) are characterized by the presence of ferromagnetic metallic regions embedded in the low conducting paramagnetic matrix above $T = T_{Curie}$, $T_{Curie}$ is the Curie temperature. At $T = T_{Curie}$ one can observe a percolative transition to the macroscopic ferromagnetic metallic state. The transition from the "pseudogap" to the macroscopic dissipationless state is also of percolative nature.

## 3.2. Phase separation

The picture described above can be treated in the framework of a general concept of phase separation. This concept was introduced by Gor'kov and Sokol (1987) shortly after the discovery of the high $T_C$ oxides (Bednorz and Mueller (1986)) and then was studied in many papers (see, e.g., book by Sigmund and Miller, Eds. (1994)). This concept implies the coexistence of metallic and insulating phases, and such a coexistence is a very important ingredient of the physics of doped cuprates. Of course, this concept means that the nominal state of the compound is in itself inhomogeneous; this feature is manifested by the separation of metallic and insulating phases.

An interesting analysis of NMR data on nuclear spin relaxation in cuprates (Gor'kov and Teitel'baum (2003)) has demonstrated that below T* the temperature dependence of $^{63}T_1^{-1}$ can be presented as a sum of two contribution:

$$^{63}T_1^{-1} = {}^{63}T_1^{-1}(x) + {}^{63}\tilde{T}_1^{-1}(T) \qquad (3.1)$$

where $T_1(x)$ is sample dependent and $\tilde{T}_1(T)$ is an universal function of temperature. The first term corresponds to the "stripe" like excitations and the second one to moving metallic and antiferromagnetic subphases.

Our picture of the "pseudogap" state (Fig. 7) implies the next step in the picture of inhomogeneity. Namely, in addition to the mixture of metallic and insulating phases, the metallic phase is itself inhomogeneous; we are dealing with the coexistence of normal and superconducting phases within the metallic phase.

One should again stress an important aspect of the inhomogeneity of the metallic phase (mixture of normal and superconducting regions, Fig. 7). Namely, one should take into account the proximity effect between the normal matrix and superconducting "islands". For example, the proximity effect determines the minimum size of a superconducting "island" which is of the order of the coherence length $\xi_0$. If the size is smaller than $\xi_0$, the superconducting state of the "islands' would be completely destroyed by the normal metallic matrix.

### 3.3. Inhomogeneity: experimental data.

The picture of inhomogeneity described above has strong experimental support. The inhomogeneous structure of the cuprates has been observed using neutron diffraction (Egami and Billingee (1996)). The underdoped compound is very inhomogeneous but becomes more homogeneous if it is doped towards the optimum level (maximum $T_C^{res}$) Bozin et al. (2000). Such a picture is totally consistent with our scenario for the "pseudogap", because, this phenomenon is very strongly manifested in the underdoped region.

Scanning tunneling microscopy (STM) provides a wealth of information about local structure. An atomic scale study using the STM technique has been performed by the J. Davis group (Cornell University). It has been demonstrated (Pan et al. (2001)) that the presence of oxygen dopants in the $Bi_2Sr_2CaCu_2O_{8+\delta}$ compound leads to inhomogeneity which is manifested by a spatial dependence of the density of states, and, correspondingly, the order parameter is inhomogeneous. Recent STM measurements (McElroy et al. (2005b)) has demonstrated that individual dopants are the cause of local disorder, and, therefore, indeed, their statistical distribution leads to observed inhomogeneity of the sample. A picture of the phase separation has been observed by Lang et al. (2002). A spatial Fourier transform method which allows one to study atom-scale modulations and the doping dependence of the nanoscale electronic structure (McEcrey et al. (2003); McElroy et al. (2005a)).

STM measurements have been performed at different locations on the surface of a BiSrCaCuO sample (at $T \cong 4.2K$). The energy gap defined as a distance between the peaks of the density of states displays a strong spatial dependence (see also Truscott (2000)). These observations provide strong experimental support for the concept of the inhomogeneity of the metallic phase.

It was stated above (Sec.3.1.) that one should distinguish the resistive ($T_c \equiv T_c^{res.}$) and magnetic transitions ($T_c^* \equiv T_c^{Meis.}$). Experimentally this has been demonstrated by dos Santos et al. (2003); the transition temperature onset, $T_c(x)$, in the $Bi_2Sr_2Ca_{1-x}Pr_xCu_2O_{8+\delta}$ (Bi 2212+Pr) compound was analyzed in detail by resistivity and magnetization measurements. The behavior of $T_{CR}(x)$ and $T_{CM}(x)$ appear to be entirely different. It has been shown that the observed depression of $T_{CR}(x)$ corresponds to a reduction of the superconducting volume fraction and the formation of superconducting clusters ,in total agreement with the picture described above (Sec.3.1).

We described in Ch. II an interesting study of the La-based compound performed using the STM technique with magnetic imaging (Fig.1) which has directly demonstrated the presence of diamagnetic "islands" embedded in a normal matrix and that demonstrates the percolative picture as
T → $T_C^{res}$.

As was indicated above, there are two possible scenarios of inhomogeneity, but the inhomogeneous distribution of pair-breakers is a dominant factor. Indeed, according to tunneling and infrared data, the energy gap is almost temperature independent; an increase in T only leads to a decrease in the depth of the dip in the density of states. The pair-breaking scenario corresponds to this result. We will discuss it in more detail below, (Sec. 4.3) Note also, that according to NMR data (Bobrov et al. (2002)), the charge distribution does not have a nanoscale variation. However, this does not exclude large scale variations (stripes or commensurate charge distribution (Haase and Slichter (2003)).

The inhomogeneity that is observed is caused by a non-uniform distribution of pair-breakers. Qualitatively, this conclusion follows naturally from the statistical nature of the doping process.

## 3.4. Energy Scales

The name "pseudogap state" reflects the presence of a gap structure above the critical temperature. One should stress however, that this single fact, namely, the presence of an energy gap does not lead to a complete understanding of the nature of the "pseudogap" state. Indeed, there could be many reasons for the appearance of a gap structure (superconducting pairing, charge density waves (CDW), spin density waves (SDW), band gap, Coulomb disorder, etc.).

Such complex systems as the cuprates might have different channels leading to the appearance of the energy gap. As a result, the presence of the energy gap structure above $T_C^{res}$ ("pseudogap" state) we believe is caused by a combination of factors. As we know, the energy gap structure could be affected by two factors: pairing and CDW instability (see, e.g., Balseiro and Falicov (1979)). For example, the presence of a chain structure in the YBCO compound leads to a CDW instability. At the same time the continuous transition of the gap structure through $T_C^{res}$ as well as diamagnetism above $T_C^{res}$ indicates that pairing is also essential.

The real picture in the cuprates is complicated and we are dealing with three different energy scales (Kresin et al. (2004)) and, correspondingly, with three characteristic temperatures (we denote them $T_C$, $T_C^*$, and $T^*$).

Highest energy scale ($T^*$). The highest energy scale, which we have labeled $T^*$ ($\geq 5.10^2 K$) corresponds to the formation of the inhomogeneity and peculiar crystal structure of the compounds. For example, for YBCO, the formation of the chains occurs at $T^*$.

An energy gap could open in the region below $T^*$. This gap is not related to the pairing, but, as was mentioned above, there are many other sources for the appearance of a gap. For example, the presence of a chain structure in YBCO is consistent with a charge density wave and, correspondingly, to a gap on part of the Fermi surface. Nesting of states might lead to a CDW instability in other compounds as well.

Another important property of the compound below T* is its intrinsic inhomogeneity ; this is due to the statistical nature of doping and is manifested in phase separation (see above). This property implies the coexistence of metallic and insulating phases. The periodic stripe structure [Bianconi (1994$_{a,b}$), Tranquada et al. (1995,1997),Zaanen(1998)] also appears below T*.

Phase separation is a key ingredient which determines T* as a corresponding onset temperature. Its value can be determined by the NMR measurements ( see above, Eq.(3.1), Gor'kov and Tetel'baum(2003)). Such a frustrated 1st order phase transition was described by Gor'kov (2001).

<u>Diamagnetic transition ($T_C^*$)</u>. If the compound is cooled down below T*, then at some characteristic temperature we have labeled $T_c^*$ ($T_c^* \approx 2.10^2 K$) one can observe a transition into the diamagnetic state.

The characteristic temperature $T_c^*$ corresponds to the appearance of superconducting regions embedded in a normal metallic matrix (Fig.1). The presence of such superconducting clusters("islands") leads to a diamagnetic moment , whereas the resistance remains finite, because of the normal matrix. As for the energy gap, coexistence of pairing and a CDW determine its existence and value below $T_C^*$. It is remarkable that the superconducting state appears at a temperature $T_c^*$ which is much higher than the resistive $T_c$. This value of $T_C^*$ corresponds to the real transition to the superconducting state (one can call it an "intrinsic critical temperature", see Kresin et al. (1996,1997).

Strictly speaking, the experimentally measured value of $T_C^*$ lies below $T_C^{intr.}$; the value of $T_C^{intr.}$ is depressed because of the proximity effect. Nevertheless, $T_C^*$ is an experimentally determined important parameter. It corresponds to the appearance of diamagnetic "islands" and reflects the impact of pairing. At the same time, the value of $T_C^*$, unlike $T_C^{res}$, is not noticeably depressed by pair breaking. The superconducting phase appears, at first, as a set of "isolated" islands.

The picture of different energy scales, $T^*$ and $T_c^*$ just described is in total agreement with interesting experimental data by Kudo et al. ( 2005a,b ). The impact of external magnetic field was studied by out-of –plane resistive measurements. According to the study, there are, indeed, two characteristic temperatures (Kudo et al. dubbed them as $T^*$ and $T^{**}$; $T^*>T^{**}$). The behavior of the resistivity appears to be independent on magnetic field in the region $T^*>T>T^{**}$, but strongly affected by the field at $T<T^{**}$. According to Kudo et al., the state formed below $T^{**}$ is related to superconductivity. The characteristic temperatures $T^*, T^{**}$ directly correspond to the energy scales $T^*$ and $T_c^*$ introduced above.

<u>Resistive transition ($T_C \equiv T_C^{res}$)</u> As the temperature is lowered below $T_c^*$, new superconducting clusters appear ( Fig.7) and existing clusters form larger " islands". This is a typical percolation scenario. At some characteristic temperature ($T_c$) the macroscopic superconducting phase is formed ("infinite" cluster in terms of the percolation theory, see, e.g., Sklovsky and Efros (1984)). The formation of a macroscopic phase at $T_c$ leads to the appearance of a dissipationless state (R=0) .

It is important also to stress, that in the region $T_c^*>T>T_c$ each "island" has its own phase, so that there is no phase coherence for the whole sample . Macroscopic phase coherence appears only below $T_c$.

Therefore, there are three different energy scales and, correspondingly, three characteristic temperatures $T^*$, $T_C^*$, $T_C$. (Fig.8).

The value of $T_C$ is lower than $T_C^*$ because of local depressions caused by the pair-breaking effect and an inhomogeneous distribution of pair-breakers (dopants). It is interesting to note that the value of $T_C^*$ is an intrinsic value of the critical temperature. This value is noticeably higher than the resistive $T_C$.

To conclude this section, let us stress again that the inhomogeneous distribution of pair-breakers (dopants) along with local depressions in the value of critical temperature leads to a spatial dependence of $T_C$ , i.e.. $T_C(r)$. (Fig. 7). The value of $T_C^*$ corresponds to an "intrinsic" $T_C^{intr.}$; this temperature corresponds to the transition into the superconducting state in the absence of pair-breaking and indeed, has a value $T_C^{intr.} \approx 210^2 K$ (Kresin et al. (1996)). It is important to note that the value of $T_C^*$ is much higher than $T_C \equiv T_C^{res}$ .

## IV.    Theory

In this section we are going to present the theoretical analysis of the main features of the pseudogap state: density of states and the appearance of a gap structure, diamagnetism, a.c. properties, and the "giant" Josephson proximity effect. The analysis is based on our model (Ovchinnikov et al. 1999, 2001, 2002), Kresin et al. 2000, 2003, 2004); the qualitative picture was described above, Ch. III.

### 4.1. General equations

Inhomogeneity of the system is a key ingredient of the theory. Because of it, it is convenient to use a formalism describing the compound in *real* space. That's why we employed the method of integrated Green's function which was developed by Eilenberger (1969) and independently by Larkin and Ovchinnikov (1968), see also the review by Larkin and Ovchinnikov (1986).

The main equations have the form:

$$\alpha\Delta - \beta\omega + \frac{D}{2}\left(\alpha\partial_-^2\beta - \beta\partial_{\vec{r}}^2\alpha\right) = \alpha\beta\Gamma \tag{4.1}$$

$$|\alpha|^2 + |\beta|^2 = 1 \tag{4.1'}$$

$$\Delta = 2\pi T|\lambda|\sum_{\omega>0}\beta \tag{4.1''}$$

Here $\alpha$ and $\beta$ are the usual and pairing Green's functions averaged over energy, $\Delta$ is the order parameter, $\Gamma \equiv \tau_s^{-1}$ is the spin-flip relaxation time. Because of the inhomogeneity, all of these quantities are spatially dependent. In addition, $\partial_\pm = \partial_{\vec{r}} \pm 2ie\vec{A}$, $\vec{A}$ is the vector potential, $\partial_{\vec{r}} = (\partial/\partial\vec{r})$. We consider the "dirty" case, so that D is the diffusion coefficient.

These equations contain the spatially dependent functions $\alpha$, $\beta$, $\Delta$. The method is very effective for treatment of spatially dependent properties.

## 4.2. Diamagnetism

The Cu-O layers contain superconducting "islands" and their presence leads to an observed diamagnetic moment. Because of the dependence $T_C(\mathbf{r})$, the size of the superconducting region occupied by the "islands" decreases as temperature is increased. As a result, diamagnetism which is strongly temperature dependent is observed.

The evaluation of the diamagnetic moment (Ovchinnikov et al. (1999)) will now be described. Based on Eqs. (4.1), one can calculate the order parameter $\Delta(\mathbf{r})$ and then the current $j(\mathbf{r})$. Then, one can calculate the magnetic moment, since the magnetic moment for an isolated cluster is

$$M_z = L \int d\rho [\rho j]_z \tag{4.2}$$

Here L is the effective thickness of the superconducting layer.

Assume that the sample contains a sufficient amount of magnetic impurities so that $\tau_s T_c^\circ \ll 1$; as a result $T_c \ll T_c^\circ$, where $T_c$ is the average value of the critical temperature, and $T_c^\circ$ corresponds to the transition temperature with no magnetic impurities. In this case, with the use of Eqs. (4.1), we obtain

$$\Delta = 2\pi T |\lambda| \sum_{\omega > 0} \left( \Gamma + \omega - \frac{D}{2} \partial_-^2 \right)^{-1} \left\{ \Delta - \frac{\omega}{2} \beta_0 |\beta_0|^2 + \frac{D}{4} \beta_0 \partial_r^2 |\beta_0|^2 \right\} \tag{4.3}$$

The order parameter can be found in the form $\Delta = C\Delta_o$ where C is a constant. As a result, we arrive at the following equation:

$$\ln(T_c^o/T) = \psi\left[1/2 + \frac{\Gamma_\infty + \lambda_1}{2\pi T}\right] - \psi(1/2) + \frac{C^2}{12\Gamma_\infty^2} \frac{\left(\Delta_0^{*2}, \Delta_0^2\right)}{\left(\Delta_0^*, \Delta_0\right)} \quad (4.4)$$

Here $\psi$ is the Euler function, and the notation (f,g) corresponds to scalar product of the functions. The transition temperature $T_C \equiv T_C^{aver}$ is determined by the equation which can be obtained from Eq. (4.4) if we insert $C = \lambda_1 = 0$:

$$\ln(T_c^o/T_c) = \psi\left[1/2 + (\Gamma_\infty/2\pi T_c)\right] - \psi(1/2) \quad (4.5)$$

which is the well-known pair-breaking equation (Abrikosov and Gor'kov (1961)). Eq. (4.4) is the generalization of Eq. (4.5) for the inhomogeneous case.

The current density is described by the expression, (Larkin and Ovchinnikov (1969)):

$$j = -ie\upsilon D\pi T \sum_\omega (\beta^* \partial_- \beta - \beta \partial_+ \beta^*) \quad (4.6)$$

Here $\upsilon$ is the density of states. With the use of Eqs. (4.4) and (4.6), we obtain:

$$j = -\frac{ie\upsilon DC^2}{2\pi T} \psi'\left(1/2 + \frac{\Gamma_\infty + \lambda_1}{2\pi T}\right)\left(\Delta_0^* \partial_- \Delta_0 - \Delta_0 \partial_+ \Delta_0^*\right) \quad (4.7)$$

where $\Delta_0$ is the solution of the equation:

$$\left(\Gamma - (D/2)\partial_-^2\right)\Delta_o = (\Gamma_\infty + \lambda_1)\Delta_o \quad (4.8)$$

As a result we can obtain the following expression for the magnetic moment of an isolated cluster $M_z = L\int d\vec{\rho}\left[\vec{\rho}\vec{j}\right]_z$:

$$M_z = -\left(e^2 \upsilon DC^2 HL/\pi T\right)\psi'\left(\frac{1}{2} + \frac{\Gamma_\infty + \lambda_1}{2\pi T}\right) K \quad (4.9)$$

Here $K = \int d\vec{\rho}\rho^2 \Delta_o^2$, and the vector-potential has been chosen as $\vec{A} = \frac{1}{2}\left[\vec{H}\vec{r}\right]$; L is the effective thickness of the superconducting layer. Note also that because the cluster size is smaller than the penetration depth, one can neglect the spatial variation of the magnetic field.

Consider the most interesting case when the variation of the amplitude $\delta\Gamma(\vec{r}) = \Gamma_\infty - \Gamma$ has the form

$$\delta\Gamma(\vec{r}) = \begin{cases} \delta\Gamma(\rho) & ; \quad \rho < \rho_o \\ 0 & ; \quad \rho > \rho_o \end{cases}$$

Then Eq. (4.8) can be written in the form

$$\left[\delta\Gamma(\rho) - \frac{D}{2}\left(\frac{1}{\rho}\frac{\partial}{\partial\rho}\left(\rho\frac{\partial}{\partial\rho}\right) - e^2 H^2 \rho^2\right)\right]\Delta_0(\rho) = \lambda_1 \Delta_0(\rho) \quad (4.10)$$

A similar equation has been studied by Ovchinnikov et al. (1996) The solution is:

$$\Delta_0(\rho) = \frac{1}{\sqrt{eH\rho^2}}\begin{cases} M_{(\lambda_1+\delta\Gamma)/2eHD;\,0}(eH\rho^2) & ; \quad \rho < \rho_0 \\ C_1 W_{\lambda_1/2eHD;\,0}(eH\rho^2) & ; \quad \rho > \rho_0 \end{cases} \quad (4.11)$$

Here $W_{\lambda,\mu}(z)$ and $M_{\lambda,\mu}(z)$ are the Whittaker functions.

Finally, one can obtain the following expression for the magnetic moment

$$M_z = -A(\tilde{B} - \tau^2)H \quad (4.12)$$

Here

$$A = (8\pi^2 e^2 \upsilon D T_C^2 / \Gamma_\infty)\rho_0^2 z_0^{-4} n(\tilde{x}_{3;2}\tilde{x}_{2;1}/\tilde{x}_{1;4}); \quad \tilde{B} = B+1 \quad (4.12')$$

$$B = -6\lambda_1\Gamma_\infty/(\pi T_C)^2$$

and

$$\lambda_1 = -\delta\Gamma + 0.5D(z_o/\rho_o)^2 \qquad (4.12")$$

n is the cluster concentration, and $\tilde{x}_{n;i} = \int_0^{z_o} dx \cdot x^n J_o^i(x)$. If $\delta\Gamma \ll \Gamma_\infty \cong \pi T_C^o$, the value of the local critical temperature $T_{C;L}$ greatly exceeds its average value; $z_0 \cong 2.4$.

One can see directly from Eq. (4.12), that it is possible to observe a noticeable diamagnetic moment. Indeed, if we assume realistic values: p=$10^{-20}$cm-sec$^{-2}$, l=40 Å (l is a mean free path: D=$v_F$l/3), $T_C$=10K, $\Gamma_\infty$=$10^2$ K, $\delta\Gamma$=50K, $\rho_0$=80 Å, and

$n \cong 0.1$, we obtain the following values of the parameters:

A$\cong 10^{-5}$, B=3, $|\lambda_1|$=5K.  Then, for example, at T=11K, one can observe $\chi_D$=$M_Z$/H=3×$10^{-5}$; this value greatly exceeds the usual paramagnetic response of a normal metal, $\chi_P \cong 10^{-6}$.

A diamagnetic response can be observed in the region $\tau < \tilde{B}$. It is important to note that the limitation on the value of $\tilde{B}$ is caused by the proximity effect. Indeed, the value of $|\lambda_1|$ is defined by equation (4.12"). The value of $|\lambda_1|$ depends on an interplay of two terms. The first term reflects the impact of the magnetic scattering, and the second negative term describes the proximity effect. For example, a decrease in the size of the inhomogeneity $\rho_0$ leads to an increase of the second term and, accordingly, to decrease in value of $\tilde{B}$, thus decreasing the temperature region ($\tau < \tilde{B}$) in which one can observe a diamagnetic response. This is natural, since the influence of the proximity effect (see, e.g., Gilabert(1977)) to depress the superconductivity grows with a decrease in the size $\rho_0$ of the superconducting grain.

As previously mentioned, Bergemann et al. (1998) described torque measurements performed on a $Tl_2Ba_2CuO_6$ overdoped sample ($T_c \cong 15K$; the value of $T_c \cong 15K$ was determined by resistive measurements). A diamagnetic moment, proportional to the external magnetic field, has been observed at $T>T_c$. An analysis based on the theory described above is in very good agreement with the data (Fig.3)

### 4.3. Density of states: gap structure

Based on Eqs. (4.1), one can evaluate the density of states for inhomogeneous system (Fig.9). As mentioned before, there are two possible scenarios for the appearance of an inhomogeneous structure: (1) an inhomogeneous distribution of pair-breakers (as we know, the presence of a pair breaker leads to a local depression in $T_C$), and (2) an inhomogeneous distribution of carriers leading to a spatial dependence of the coupling constant $\lambda \equiv \lambda(\mathbf{r})$.

Let us focus on the first scenario, since the inhomogeneous distribution of pair-breakers appears to be a major factor which determines the spatial dependence of the temperature $T_C^{res} \equiv T_C^{res}(\mathbf{r})$.

The energy gap manifests itself as a dip in the low frequency region of the density of states $\nu \equiv \nu(\omega)$. The density of states is defined by the relation $\nu = \text{Re}\alpha$, where $\alpha$ is the usual Green's function averaged over energy (see Eqs. (4.1))

The calculation of α (Ovchinnikov et al. (2001)) leads to the following expression for its average value:

$$\langle \alpha \rangle = 1 - \left(n_c C^2 / 2\right)\left[\lambda^2 - \omega^2 + 2i\lambda\omega\right]\left(\lambda^2 + \omega^2\right)^{-2} \qquad (4.13)$$

where

$$C^2(T) = A(T_C^L / T) + \Psi(0.5 + \lambda / 2\pi T_C^L) \quad - \Psi(0.5 + \lambda / 2\pi T)$$
$$A^{-1}(T) = (4\pi T)^{-1}\left[\Psi'(0.5 + \lambda / 2\pi T) + (\lambda / 4\pi T)\right.$$
$$\left.\times \Psi''(0.5 + \lambda / 2\pi T)\right] - \left(D / 2(2\pi T)^2\right) \qquad (4.13')$$
$$\times \Psi''(0.5 + \lambda / 2\pi T) \int d\vec{\rho}\Delta_0^2 \left(\partial \Delta_0 / \partial \rho\right)^2$$

where $\lambda = \lambda_1 + \Gamma_\infty$, and the eigenvalue $\lambda_1$ is determined by Eq.(4.12'')

The density of states for the inhomogeneous system of interest is plotted in Fig.9. One can see directly that, indeed, there is a "softening" of the low-energy part of the density of states, and this is a clear manifestation of the "pseudogap" structure.

If the temperature is above $T_C^{res}$ and is increased towards $T_C^*$ then the difference $\Delta\upsilon = \upsilon_{max} - \upsilon_{min} \to 0$. At the same time the position of the peak is independent of T. This feature is very specific for the "pseudogap" phenomenon caused by an inhomogeneous distribution of pair breakers.

Indeed the density of states and its temperature dependence were directly measured by tunneling spectroscopy by Renner et al. (1998). One can see directly from the data (see Fig.4), that the gap structure ("pseudogap") persists above $T_C$, but the peak position does not depend on temperature. This is in an agreement with the scenario discussed above and, therefore, for the $Bi_2Sr_2CaCu_2O_{8+\delta}$ sample studied by Renner et al. (1998) the "pseudogap" phenomenon is caused by an inhomogeneous distribution of pair breakers.

An interesting measurement of the interlayer tunneling spectroscopy for an overdoped $Bi_2Sr_2CaCu_2O_{8+\delta}$ compound was described by Suzuki et al. (1999). The authors also observed the occurrence of the pseudogap below 150K, that is at much higher temperatures than the resistive $T_C \cong 87K$. The data are also consistent with the picture of an inhomogeneous distribution of pair breakers.

Infrared spectroscopy data discussed above (see Sec.2.4) also is in agreement with the conclusion that the inhomogeneous distribution of pair-breakers is a major source of the inhomogeneity in the cuprates.

Generally speaking, an inhomogeneous charge distribution and, correspondingly, the dependence $\lambda(\mathbf{r})$ can also lead to a spatially dependent $T_C$. However, one can show that in this case the value of the gap decreases with increasing T in the region above $T_C^{res}$ (Ovchinnikov et al. (2001)). Therefore, based on various experimental data, one can conclude that the inhomogeneous distribution of pair-breakers caused by the statistical nature of the doping, plays a dominant role as a source of inhomogeneity.

**4.4. ac transport**

The dc transport properties of the inhomogeneous system above $T_C^{res}$ is determined by the normal phase, since only this phase can provide a continuous path. The situation with ac transport is entirely different, and the superconducting "islands" make a direct contribution to the ac conductivity and to surface impedance.

As we know, the real and imaginary parts of the surface impedance of a normal metal are almost equal (see e.g., Landau and Lifshitz, 1960). The situation is entirely different in superconductors (see e.g., Tinkham (1996). To describe the "pseudogap" state, it is interesting to consider an inhomogeneous system which consists of normal and superconducting regions (Ovchinnikov and Kresin, 2002)

The analysis is based on Eqs. (4.1); the system is in an external field

$$\mathbf{A}(\mathbf{r}, \tau) = \exp(-i\omega_0 \tau)\mathbf{A}(\mathbf{r}); \quad \varphi(\mathbf{r}, \tau) = \exp(-i\omega_0 \tau)\varphi(\mathbf{r})$$

($\mathbf{A}$ and $\varphi$ are the vector and scalar potentials, $\omega_0$ is the frequency, $\tau$ is an imaginary time). The formalism of thermodynamic Green's functions is employed (see, e.g., Abrikosov et al. (1963))

One can formulate a complete general system of equations determining the ac response for an inhomogeneous system. Based on these equations, one can calculate the ac conductivity, and then the surface impedance Z, since

$$Z = \left(\frac{\omega}{4\pi\sigma}\right)^{1/2} \exp(-i\pi/4) \qquad (4.14)$$

(see, e.g. Landau and Lifshitz (1960)) For normal metals the difference between ReZ and ImZ is negligibly small and is connected with the dependence:

$\sigma(\omega) = \sigma_0 (1 - i\omega\tau_{tr})^{-1}$; in our case $\omega\tau_{tr} \ll 1$. The situation for the "pseudogap state

is different. One can show that a metallic compound which contains superconducting "islands", is characterized by a strong inequality: ReZ ≠ |Im Z|.

In the temperature region close to $T_C^*$, i.e. $(T-T_C^*) \ll T_C^*$, the expressions for the conductivity $\sigma_{eff}$ and, correspondingly, for the impedance, can be simplified and has the form:

$$\text{Re } Z = \tilde{Z}_n \left[1 - (\sigma_2/2\sigma_1)\right]$$
$$\text{Im } Z = -\tilde{Z}_n \left[1 + (\sigma_2/2\sigma_1)\right] \qquad (4.15)$$

Here

$$\sigma_2/2\sigma_1 = (s/\omega)(T_C^* - T) \qquad (4.15')$$

or

$$\text{Re } Z = -\text{Im } Z - \omega^{-1}(2s\tilde{Z})(T_c^* - T) \qquad (4.16)$$

Here $\tilde{Z}_n = (\omega/8\pi\sigma_1)^{1/2}$, $S \equiv S(n_s)$. the quantity s depends exponentially on $n_S$. Note that ReZ = |ImZ| at $T \geq T_C^*$. This can be seen directly from (4.15), (4.16). The inequality ReZ ≠ |ImZ| at $T \leq T_C^*$ is caused by the presence of superconducting "islands" and is described by the second term in Eq.(4.16). It is important to note that this term is proportional to $\omega^{-1/2}$. A relatively small value of the frequency $\omega$, e.g. in the microwave region, leads to a noticeable contribution to the impedance. In addition, the dependence $\propto \omega^{-1/2}$ can be directly measured experimentally.

Eqs (4.15) and (4.16) describe the ac response of an inhomogeneous superconductor in the pseudogap region ($T_C^{res} < T < T_C^*$).

The quantity s depends on a number of experimentally accessible parameters, including geometry. For example, if we assume the values:

$$T_C^* = 200K, \quad T_C^{res} = 110K; \quad \Gamma_\infty = 160K,$$

$$a = 2.5\xi; \quad \xi = (D/T_C)^{1/2}, \quad n_S \cong 10^{-2}, \quad \omega = 2\pi 10^{10} \sec^{-1}, \quad (4.17)$$

we obtain:

$$s \cong 4 \bullet 10^8 s^{-1} K^{-1} \quad (4.18)$$

Measurements of Z for the $HgBa_2Ca_2Cu_3O_{8-\delta}$ compound at $T > T_C^{res}$ were performed by Kusko et al. (2002). It has indeed been observed that the slopes of the temperature dependencies are different meaning that $ReZ \neq ImZ$. Using experimentally determined value of the slope, one can calculate the parameter s, and it is close to the value we estimated above(4.18). This substantiates the choice of parameters we used.

Let us make several comments related to the d.c. transport. As we mentioned above, (Ch. II), the normal resistivity above $T_C$ displays a specific non-linear behavior. Such an effect can be understood as a result of the interplay between normal matrix and superconducting "islands". A decrease in temperature leads to an increase in the superconducting fraction, and this affects the total normal resistivity. A relevant analysis of the normal resistivity for the $Bi_2Sr_2Ca_{1-x}Cu_2O_{8+\delta}$ compound, based on the inhomogeneous picture, has been carried out by dos Sandos et al. (2003). An interesting study of the in-plane transport for manganite – YBCO heterostructures (Soltan et al. (2005)) has shown that the in-plane resistance drops as a result of the spin-injection into the Cu-O plane in the c-direction. This effect can be explained by an increase in the number of pair-breakers caused by the injection; this leads to an increase in the "normal" fraction. The impact of the "pseudogap" on thermal conductivity (Minami et al. (2003)) has demonstrated that the "pseudogap" phenomenon does not arise from fluctuation. As a whole, one should note that transport phenomenon deserves additional theoretical and experimental study.

**4.5. Isotope effect**

The isotope effect on $T_C^*$ observed experimentally (Sec. 2.6) also reflects the fact that the superconducting pairing persists above the resistive transition. It is interesting to note that the isotope coefficient has a negative sign. This unusual feature is consistent with the microscopic model of the isotope effect in the high $T_C$ oxides (Kresin and Wolf (1994)). Indeed, a strong non-adiabaticity (axial oxygen in YBCO is in such state) results in a peculiar polaronic isotope effect.

Namely, the doping, and therefore, the carrier concentration n, and, correspondingly, $T_c$ are affected by the isotope substitution. If the charge transfer occurs in the framework of the usual adiabatic picture, so that only the carrier motion is involved, then the isotope substitution does not affect the forces and therefore does not change the charge transfer dynamics. However, the situation of strong non-adiabaticity is different and does not allow the separation of electronic and nuclear motion; in this case charge transfer appears as a more complex phenomenon which does involve nuclear motion, and this leads to a dependence of the doping on isotopic mass.

Consider the case when the lattice configuration corresponds to the situation when some definite ion (e.g., axial oxygen for YBCO) is in (or near) a degenerate state; this means that the degree of freedom describing its motion corresponds to electronic terms crossing (see Fig.10). Then the ion has two close equilibrium positions (double-well structure (Fig10)), Then it is convenient to use a so-called "diabatic representation" (see, e.g., O'Malley (1967), Smith (1969), Kresin and Lester (1984), Dateo et al. (1987)). In this representation we are dealing directly with the crossing of electronic terms. The operator $H_{el.} = \hat{T}_{\vec{r}} + V(\vec{r}, \vec{R})$ ( $\hat{T}_{\vec{r}}$ is a kinetic energy operator, $V(\vec{r}, \vec{R})$ is a total potential energy, $\vec{r}$ and $\vec{R}$ are the electronic and nuclear coordinates, correspondingly) has non-diagonal terms (unlike the usual adiabatic picture when $\hat{H}_{el}$ is diagonal). The charge transfer in this picture is accompanied by the transition to another electronic term. Such a process is analogous to the Landau-Zener effect.

Consider the axial oxygen in YBCO. Its dynamics is described by the double-well structure, and such a structure has been observed experimentally with use of the x-ray absorption fine structure technique (Haskel et al. (1997)), Fig. 11. The charge transfer in this case is described by polaronic motion (dynamic polaron). Note that a similar effect leads to the isotope effect in manganites (see Gor'kov and Kresin (2004)).

The total wave function can be written in the form

$$\Psi(\vec{r}, \vec{R}, t) = a(t)\,\Psi_1(\vec{r}, \vec{R}) + b(t)\,\Psi_2(\vec{r}, \vec{R}) \qquad (4.19)$$

Here

$$\Psi_i(\vec{r}, \vec{R}) = \psi_i(\vec{r}, \vec{R})\,\Phi_i(\vec{R}), \quad i = \{1, 2\}$$

$\psi_i(\vec{r}, \vec{R})$, $\Phi_i(R)$ are the electronic and vibrational wave functions that correspond to two different electronic terms (see Fig.10).

In the diabatic representation the transition between the terms are described by the matrix element $V_{12}$, where $V \equiv \hat{H}_{\vec{r}}$. One can show that

$$V_{12} \cong L_0\,F_{12} \qquad (4.20)$$

where

$$L_o = \int d\vec{r}\,\psi_2^*(\vec{r},\vec{R})\,\hat{H}_r\,\psi_1(\vec{r}, \vec{R})\,\Big|_{R_o} \qquad (4.20')$$

is the electronic constant ($R_0$ correspond to the crossing configuration), and

$$F_{12} = \int \varphi_2^*(\vec{R})\varphi_1(\vec{R})dZ \qquad (4.20'')$$

is the Franck-Condon factor. The presence of the Franck-Condon factor is a key ingredient of our analysis. Its value strongly depends on the ionic mass and, therefore is affected by the isotope substitution.

The calculation (Kresin and Wolf (1993)) leads to the following expression for the isotope coefficient:

$$\alpha = \gamma \frac{n}{T_c} \frac{\partial T_c}{\partial n} \qquad (4.21)$$

where γ has a weak logarithmic dependence on ionic mass M.

Based on Eq. (4.21), one can explain why the isotope coefficient has a negative value. It is important to note that $\alpha \propto \partial T_c / \partial n$. Indeed, for the "pseudogap" state it means that $\alpha \propto \partial T_c^* / \partial n$, since $T_C^*$ is the "intrinsic" value of the critical temperature. The isotope coefficient α is a negative, because $(\partial T_c^* / \partial n) < 0$. Indeed, increase in doping in the underdoped region leads to decrease in the value of $T_C^*$ (at optimum doping $T_C^* \cong T_C$), and this is due to an increase in a number of dopants that is, the pair-breakers.

## V. "Giant" Josephson proximity effect

We mentioned above (see Ch II) an interesting experimental study of S-N-S Josephson junctions (Bozovic et al. (2004)). A finite Josephson current was observed for junctions with $L \gg \xi_n$ (L is the thickness of the N-barrier and $\xi_n$ is the proximity coherence length). This phenomenon can not be explained by the usual theory of S-N-S proximity junctions.

We focus on the especially interesting case of $S-N'-S$ (for a general discussion see review by Devin and Kleinsasser (1996)) junctions where the electrodes are the high Tc superconducting films ( e.g., $La_{0.85}Sr_{0.15}CuO_4$, or $YBa_2Cu_3O_7$), and the barrier $N'$ is made of the underdoped cuprate, so that $T'_c < T_c$, $T_C' \equiv T_C^{N'}$. The generally accepted notation $N'$ emphasizes a difference between $SN'S$ and a typical SNS junction. Here we consider temperatures when the barrier is in the normal resistive state between $T_c^S$ and $T'_c$. The use of the underdoped cuprates is useful for various device applications because the structural similarities between the electrodes S and the barrier $N'$ eliminates many interface problems. The "giant" phenomenon is manifested in a finite superconducting current through the $S-N'-S$ Josephson junction with a thick $N'$ barrier, so that

$L \gg \xi_N$ ( L is the thickness of the barrier, $\xi_N$ is the proximity coherence length). The configuration such that the layers forming the barrier N' are parallel to the electrodes so that the Josephson current flows in the c-direction . Then the coherence length is very short $\xi_c \approx 4A$, so that we are dealing with the "clean" limit. This type of junction using the LaSrCuO material was studied by Bozovic et al. (2004) The films of $La_{0.85}Sr_{0.15}CuO_4$ ($T_C \approx 45K$) were used as electrodes, whereas the underdoped LaCuO compound ($T'_c \approx 25K$) formed the barrier. The atomic-layer-by-layer molecular beam epitaxy technique was used for these junctions and it provides atomically smooth interfaces. The barrier was prepared in the c-axis geometry. As was noted above, the coherence length $\xi_C \approx$ 4Å. The measurements were performed at $T'_c < T < 35K$. The Josephson current was observed for thickness of L up to 200Å(!). Such a "giant" effect can not be explained with use of the conventional theory. Indeed, as we know,(see ,e.g., Barone and Paterno (1982)) the amplitude of the Josephson current for the *"clean" limit* is

$$j_m = j_o \exp(-L/\xi_N) \qquad (5.1)$$

The thickness of the barrier L should be comparable with the barrier coherence length $\xi_N$, and this condition is satisfied for conventional Josephson junctions. The picture described above for the *SN'S* junctions with the cuprates is entirely different, since $L \gg \xi_N$. The superconducting current in the c-direction occurs via an intrinsic Josephson effect between the neighboring layers (see Kleiner et al. (1992)). If the barrier contains several homogeneous normal layers, then the Josephson current through such a barrier can not flow.

To understand the nature of the "giant" Josephson proximity effect it is very important to stress that the barriers we are considering are formed by underdoped cuprates. As a result, the barriers are not in the usual normal state, but in the "pseudogap" state; indeed, $T_C' < T < T_C^*$. For example, the study (Igushi et al. (2002))of the compound $La_{2-x}Sr_xCuO_4$ ( $x \cong 0.1$; $T_c \cong 20K$) in which the stoichiometry is close to that for the sample used by Bozovic et al. (2004) as the barrier. According to Igushi et al., this compound has a value of $T_C^* \cong 80K$ whereas $T_C' = T_C^{res} \cong 20K$ describes the resistive transition to the dissipationless state. The diamagnetic moment measured by Igushi et al. (2002) persists up to $T_C^*$, therefore, the question of the origin of the "giant" proximity effect is directly related to the general problem of the nature of the "pseudogap" state and this makes the question of the nature of the "giant" Josephson proximity effect particularly interesting.

This effect can be explained (Kresin et al. (2003)) by the approach described here and based on the intrinsically inhomogeneous structure of the compound. According to our model, the CuO layers forming the N-barrier contain superconducting "islands", and these "islands" form the path for the Josephson tunneling current.

For typical SNS junctions the propagation of a Josephson current requires the overlap of the pairing functions $F_L$ and $F_R$ (see, e.g., Kresin (1986)); $F_R$ and $F_L$ are pairing Gor'kov functions for left and right-side electrodes). This overlap is caused by the penetration of $F_L$ and $F_R$ ("proximity") to the N-barrier. For the system of interest here the situation is quite different. Each "island" has its own pairing function with its own phase. As a result, the Josephson current is caused by the overlap of $F_L$ and $F_1$, $F_1$ and $F_2$, etc, where $F_1$ corresponds to the "island" located at the layer nearest to the left electrode, etc. The superconducting "islands" form the network with the path for the superconducting current.

The propagation of a Josephson current through the $S-N'-S$ junction requires the formation of a channel between the electrodes. The transport of the charge in superconducting cuprates in the c-direction is provided by the interlayer Josephson tunneling (intrinsic Josephson effect see, e.g., Kleiner et al. (1992), Scheekga et al. (1998)). Therefore, the Josephson current through the barrier is measurable because of the superconducting state present in the layers.

The transfer of the Josephson current in our model implies that the electrons tunnel inside of the layers between the superconducting "islands" until one of them appears to be close to some "island" in the neighboring layer. Then the next step, namely the interlayer charge transfer via the intrinsic Josephson effect occurs, etc. As a result, the chain formed by the superconducting "islands" provides the Josephson tunneling between the electrodes and the path represents a sequence of superconducting links. It is important to note that the amplitude of the total current is determined by the "weakest" link in the chain.

The density of the critical current is determined by the equation:

$$j = A \int \exp(-r/\xi) dP \qquad (5.2)$$

or

$$j = (A/\xi) \int_0^\infty dR P \exp(-R/\xi)$$

here $\xi \equiv \xi_{11}$ is the in-plane coherence length, R is the distance between the "islands' on the same layer, $A \propto C j_{C\perp}$, C≡C(T) is the concentration of the superconducting region, so that $S_{\text{sup.}} = CS$ is the area occupied by the superconducting phase, S is the total area of the layer, $j_{C\perp}$ is the amplitude of the Josephson interlayer transition, and P is a probability of formation of chain with length R for the links, so that P=$p^{N-1}$ (see,e.g., Stuffer and Aharony (1982)), N is the number of layers forming the barrier ,and p is the probability for two neighboring in-plane "islands' to be separated by distance

r< R.

Assume that p is described by a Gaussian distribution, that is:

$$p = (c/\pi\delta)^{1/2} \xi^{-2} \int_0^R dr\, r \exp\{-\delta^{-1}[(r/\xi) - C^{-1/2}]^2\} \quad (5.3)$$

then the integral (5.3) can be calculated by the method of

steepest descent, and we obtain

$$j_m = \tilde{A} f(N) \exp(-1/C^{1/2}) \qquad (5.4)$$

Eq.(5.4) can be written in the form :

$$j_m = j_0(-1/\rho(T)) \qquad (5.5)$$

Here

$$j_0 = j_m(T_C^{res})\rho(T) = C^{1/2}(1 - C^{1/2})^{-1}$$
$$f(N) \approx \exp[-\delta \ln^{1/2}(N\tilde{l})] \quad \tilde{l} = (\pi\delta)^{-1/2}$$

We assumed that $\delta \ll 1$, $N \gg 1$.

One can see directly from Eqs.(5.4),(5.5) that the current amplitude depends strongly on temperature and it is determined mainly, by the dependence of the area occupied by the superconducting "islands" on T: $C \equiv C(T)$. In addition, there is a weak dependence of $j_m$ on the barrier thickness.

The dependence C(T) is different for various systems and is determined by the function $T_C(r)$, that is, by the nature of the doping. Note that near $T_C^*$ the value of C (T) is very small and the current amplitude is negligibly small. However, the situation is different in the intermediate temperature region and in the region $T \ll T_C^*$ which is not far from $T_C$. This is true for the data by Bozovic et al. (2004) where $T_C' = 25K$ and $T_C' < T < 35K$. For example, the value T=30K is relatively close to $T_C'$ but much below T*=80-100K. At T=30K there are many superconducting "islands", so that the value of C(T) is relatively large. At temperatures close to $T_c^{res.}$ one can use Eq. (5.5) with $\rho = a(t-1)^\nu$; t=T/$T_c$ and a=const (we have chosen a=10). One can see (Fig. 12) that such a dependence with $\nu$ =1.3 is in a good agreement with the experimental data. Note that this value for $\nu$ is close to the value of the critical index for the correlation radius in the percolation theory (see, e.g., Sklovsky and Efros (1984)).

In principle, one can use a junction with a barrier grown in the ab direction, so that the c-axis is parallel to the S electrodes. Then the path contains SNS junctions formed by the "islands" with metallic N barriers. Since $\xi_{ab} >> \xi_C$ ($\xi_{ab} \cong 20-30$Å), one should expect even a larger scale for the "giant" Josephson proximity effect with thickness L up to $10^3$Å.

Therefore, the "giant" Josephson proximity effect is also caused by intrinsic inhomogeneity of the cuprates. The "giant" scale of the phenomenon is provided by the presence of "superconducting" islands embedded in the metallic matrix and forming the chain transferring the current. The use of superconductors in the "pseudogap" state as barriers represents an interesting opportunity for "tuning" the Josephson junction on a "giant" scale.

## VI. Other Systems.

Intrinsic inhomogeneity is an essential feature of the high $T_C$ oxides, and this feature is manifested in a peculiar "pseudogap" behavior. However, the scenario is a more general one and the "pseudogap" state can be observed in other inhomogeneous systems as well. Let us describe some of these systems.

### 6.1. Borocarbides

Borocarbides represent an interesting family of novel superconductors, because they allow us to study an interesting interplay between superconductivity and magnetism (see, e.g., Canfield et al. (1998), Schmiedeshoff et al. (2001). According to data by Lascialfari et al. (2003), the borocarbides $YNi_2B_2C$ displays precursor diamagnetism above $T_C$ ($T_C = 15.25K$). Analysis of magnetization data taken with a high resolution SQUID magnetometer led to the conclusion that the unusual results were caused by an inhomogeneity of the compound and by the presence of superconducting droplets with a local value of $T_C$ higher than $T_C^{res}$. It was shown that fluctuating magnetization can not lead to the observed dependence. The presence of the superconducting isolated droplets is due to an inhomogeneity described above. Indeed, the pair-breaking effect in borocarbides as a major source of inhomogeneity (see Sec. II) was analyzed by Ovchinnikov and Kresin (2000). An unconventional temperature dependence of the critical fields Hc and $Hc_2$ observed experimentally (Schmiedeshoff et al., (2001)) was explained by the presence of pair-breakers. Pair-breaking also leads to depression of $T_C$. Statistical distribution of pair-breakers leads to spatial dependence of the critical temperature: $T_C \equiv T_C(r)$ and to the inhomogeneous "pseudogap" picture (Fig. 7).

**6.2. WO$_3$ +Na compound**

Another complex system, the Na-doped WO$_3$ compound has been studied by Reich and Tsabba (1999) and later by Shengelaya et al. (1999). The material displays a small diamagnetic moment and a concomitant decrease in resistivity. STM spectroscopy of this material (Reich et al. ) has revealed a dip in the density of states, that is, a gap structure. Probably, this system is inhomogeneous and contains superconducting "islands".

**6.3. Granular superconductors; Pb+Ag system.**

Granular superconducting films have been studied intensively before the discovery of high T$_C$ superconductivity (see, e.g., Simon et al. (1987), Dynes and Garno (1981)). These films also represent inhomogeneous superconducting systems. Such inhomogeneous films could display diamagnetic moment above T$_C$. It would be interesting to carry out a study of their magnetic properties.

An interesting example of an inhomogeneous conventional superconducting system was described by Elzinga and Uher (1985). They studied a Sn-doped Bi sample. The sample has Sn-grains embedded in a semimetallic matrix which provides the proximity charge transfer.

An interesting study of the Pb + Ag system was described recently by Merchant et al, (2001).

An electrically discontinuous (insulating ) Pb film was covered with increasing thickness of Ag (Fig.13). The Ag act to couple the superconducting Pb grains via the proximity effect. The resistive transition, as well as tunneling spectra, has been taken on a series of these films. The most insulating film has no resistive transition but a full Pb gap as revealed by the tunneling spectra. This gap is reduced as silver is added reflecting the decrease in the mean-field $T_C$ of the Pb grains. At some point, the composite film becomes continuous and superconducting with a low resistive transition temperature. The evolution of the mean-field transition temperature and the resistive transition temperature with increasing Ag thickness mimics the phase diagram of the cuprates with doping. The mean-field transition temperature resembles the pseudogap onset temperature and the resistive $T_C$ resembles the superconducting transition temperature, with the mean-field transition temperature lying above the resistive transition.

We believe that the results of the Pb/Ag artificial inhomogeneous superconductor model the behavior of the cuprates. The cuprates are doped substitutionally and inhomogenously. At some concentration of doping there are regions with a high enough concentration of carriers to locally superconduct and therefore reduce the low energy density of states. The evolution of these islands into a percolating dissipationless state would resemble the percolating proximity coupling described above. It is not then surprising that the phase diagrams would be nearly identical.

## VII. Conclusion

Inhomogeneity is an important feature of novel superconducting systems and, first of all, of high $T_C$ cuprates. This property is caused by the statistical nature of the doping combined with pair-breaking effect. As a result, the critical temperature is spatially dependent: $T_C=T_C(\mathbf{r})$. It is important to note that the "intrinsic' critical temperature ($T_C^*$) which corresponds to the formation of Cooper pairs inside of the diamagnetic "islands" (Fig.9) is much higher that the temperature of the transition into the macroscopic dissipationless and coherent state (at $T_C \equiv T_C^{res}$; "resistive" transition).

Until recently, the presence of inhomogeneites was considered as a signature of a poor quality sample (except for the " pinning" problem). However, we think that the situation is similar to that in the history of semiconductors. Indeed, initially the presence of impurities in these materials was considered as a negative factor (they were called "dirty" semiconductors). But later, when scientists developed tools allowing the precise control of the impact of various impurities (donors and acceptors), it became clear that the presence of impurity atoms is a critical ingredient; even the language had changed and sounds more "respectful" ("doped" semiconductors)The analogy between inhomogeneous novel superconductors and semiconductors is even stronger , because we are dealing with doping for both classes of materials.

The "pseudogap" state is intrinsically inhomogeneous. Since a number of superconducting properties (diamagnetism, a.c. response, etc.) persists above $T_C \equiv T_C^{res}$ up to $T_C^*$, one might think about interesting applications at temperatures higher than $T_C$. Microwave properties, Josephson effects with thick barriers ("Giant" effect, Ch. V) are especially promising. In connection with this it would be interesting to carry out more detailed experimental and theoretical studies of transport properties (especially thermal transport), a.c. response, etc. One can expect many new and unexpected results.

Figure Captions

Fig. 1.    Development of magnetic "islands" with temperature (magnetic imaging of LSCO films, Igushi et al. (2001)

Fig. 2.    Fractional area of magnetic domains with temperature for LSCO films

Fig. 3.    Diamagnetic susceptibility for the $Tl_2Ba_2CuO_{6+\delta}$ ($T_C$ = 15K). experimental data (Bergemann et al. (1998); solid line-theory (Sec. 4.2)

Fig. 4.    Tunneling conductance of states for the underdoped Bi 2212 crystal

Fig. 5.    The optical conductivity of $Yba_2Cu_3O_{6.7}$ along the c-axis

Fig. 6.    Temperature dependence of the photoemission spectra at the ($\pi$, 0) point

Fig, 7.    Inhomogeneous structure. "Islands" are characterized by values of $T_C$'s higher than the matrix

Fig. 8.    Energy scales

Fig. 9.    The behavior of the density of states for different types of inhomogeneites. The dashed line corresponds to higher temperatures.
    (a) Density of states for an inhomogeneous distribution of pair breakers;  (b) Density of states for an inhomogeneous distribution of carriers and, correspondingly, the coupling constants

Fig. 10.    Electronic terms

Fig. 11.    Occupancy of the O(2) apical oxygen

Fig. 12.   Dependence of the Josephson current on temperature: doted line – experimental data,(Bozovic et al. (2004)) solid line -theory

Fig. 13.   Pb/Ag proximity system

**Table I**

|  | Superconducting state($T<T_c$) | Normal state (($T>T_c$) | "Pseudogap" state($T_c^*>T>T_c$) |
|---|---|---|---|
| Resistance | R=0 | R≠0 | R≠0 |
| Energy gap | $\Delta \neq 0$ | $\Delta = 0$ | $\Delta \neq 0$ |
| Anomalous diamagnetism | Yes | No | Yes |
| Macroscopic phase coherence | Yes | No | No |
| Josephson effect | Yes | No | "Giant" effect |
| Isotope effect | Yes | No | Yes |
| Impedance Z | ReZ≠ImZ | ReZ=ImZ | ReZ≠ImZ |

Figure 1.

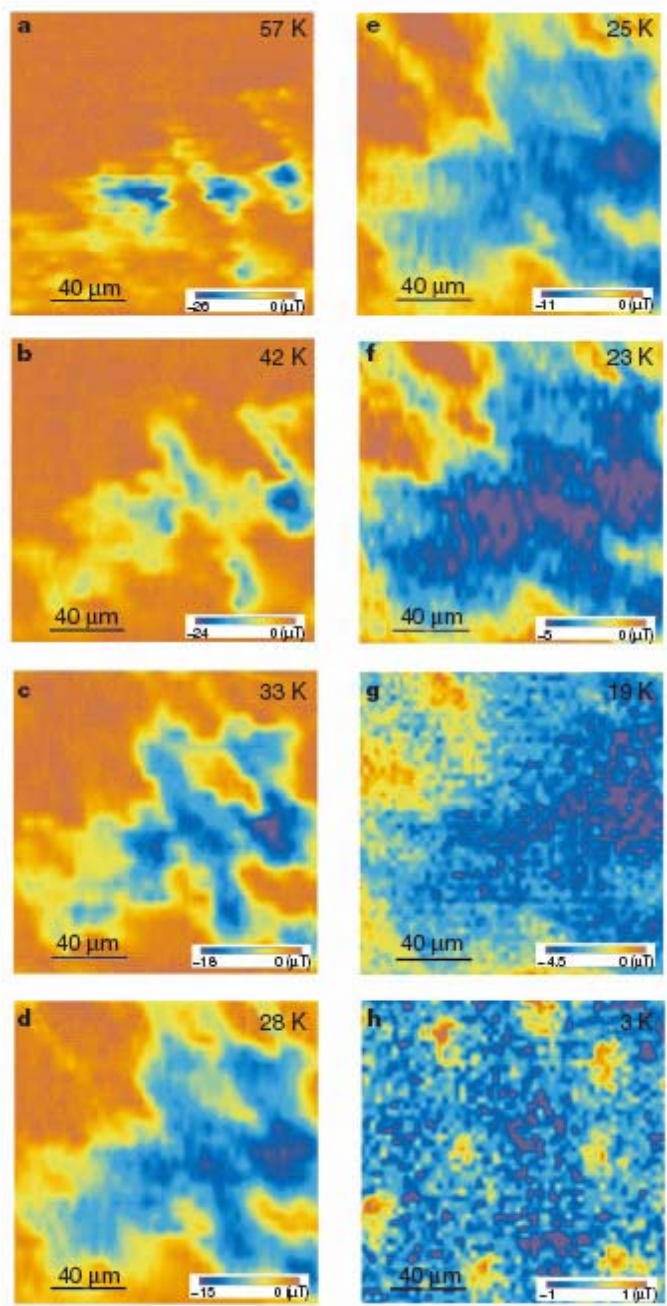

Figure 2.

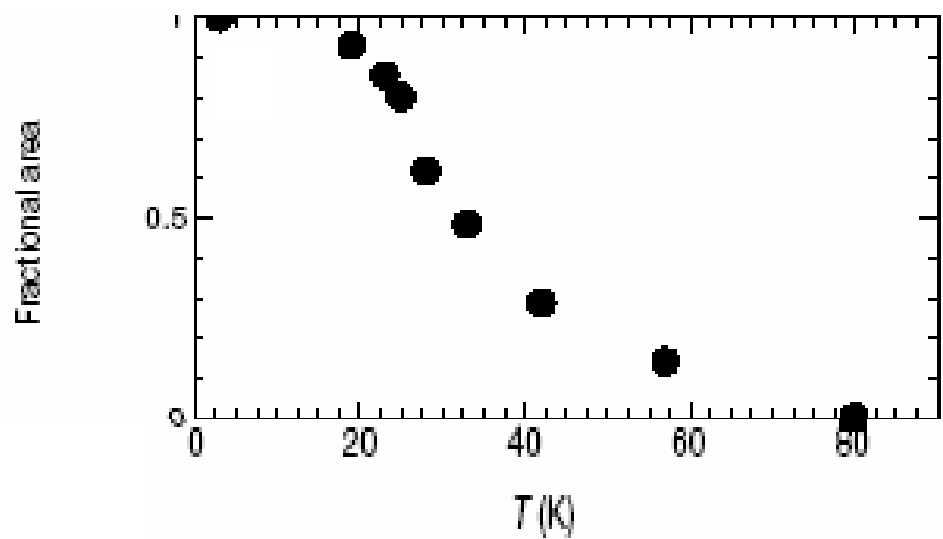

Figure 3.

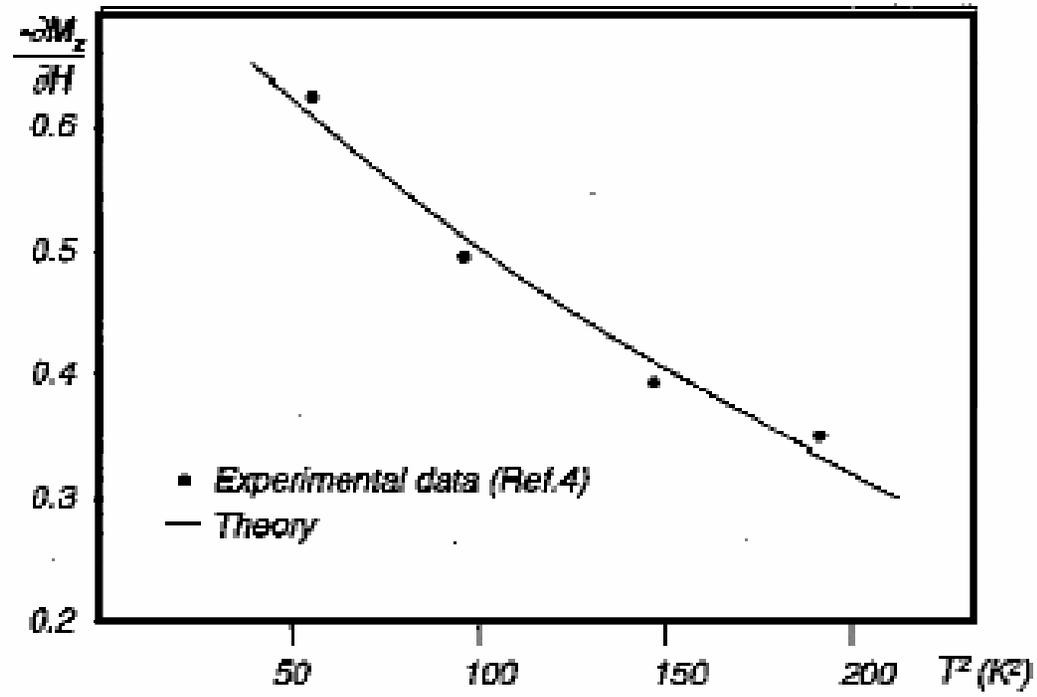

Figure 4.

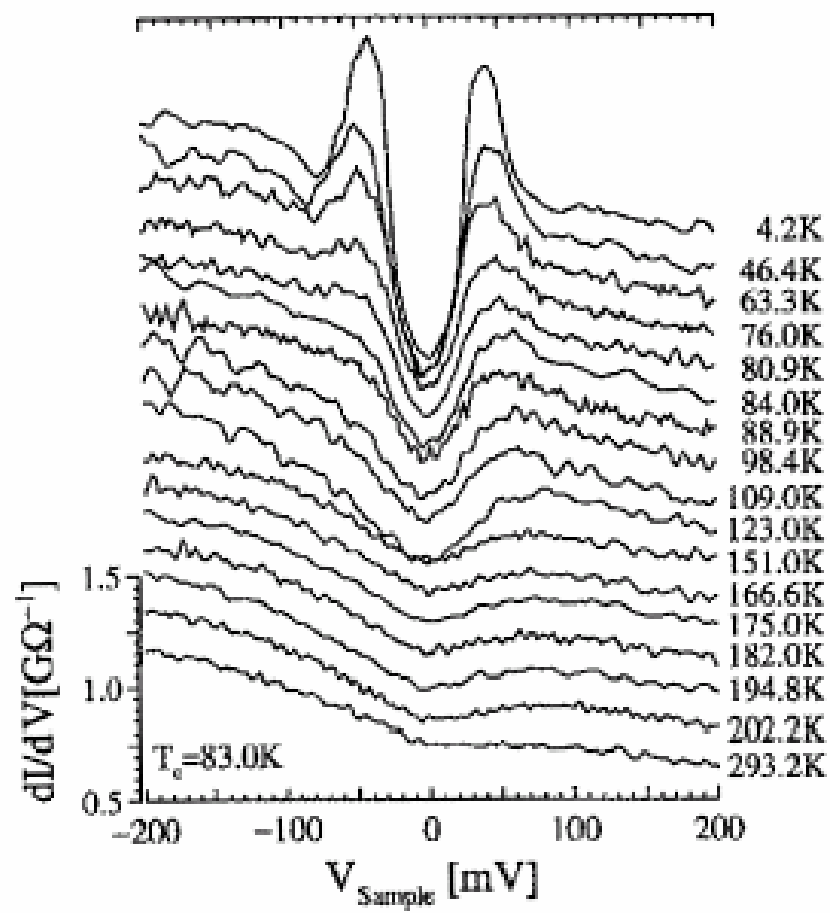

Figure 5.

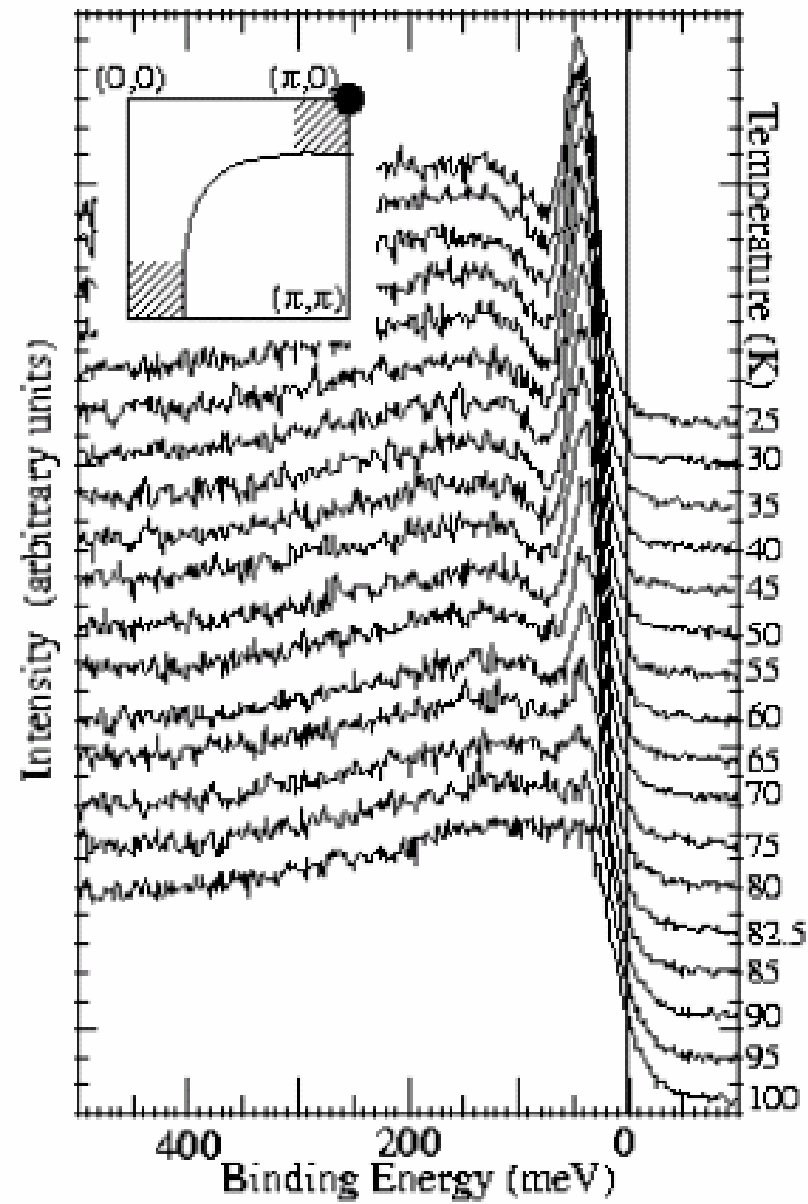

Figure 6.

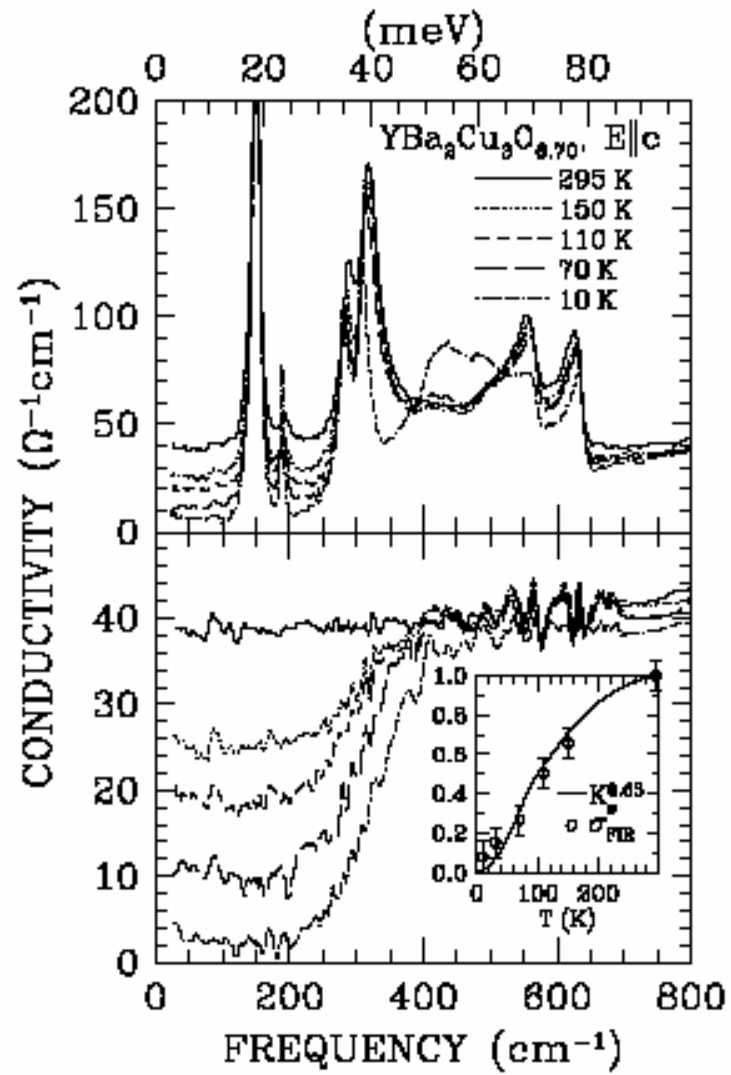

Figure 7.

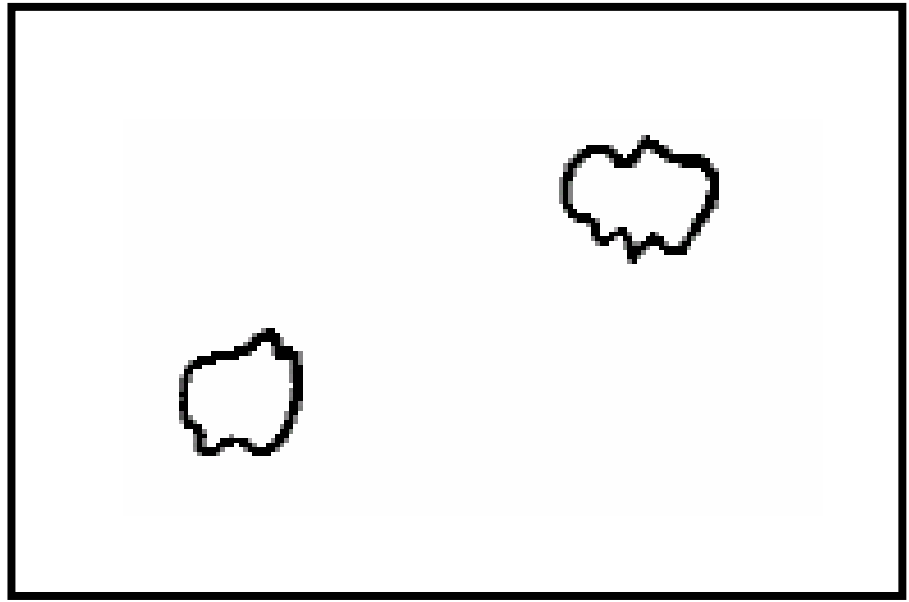

Figure 8.

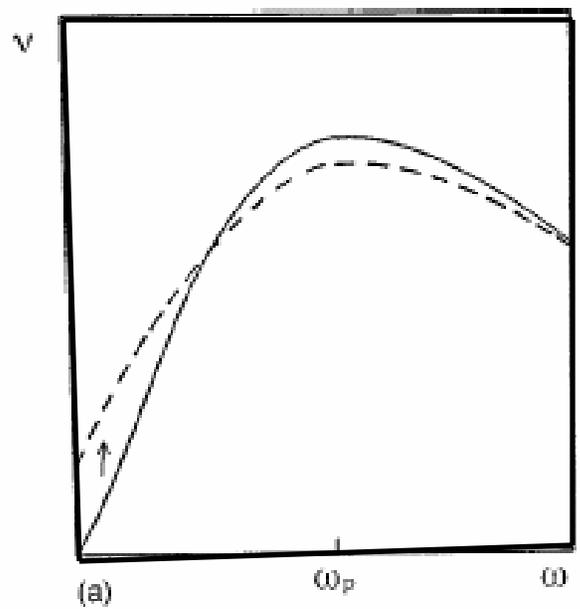

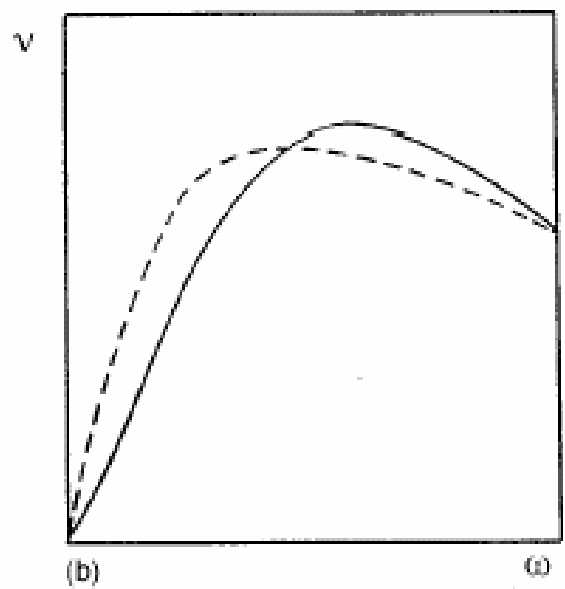

Figure 9.

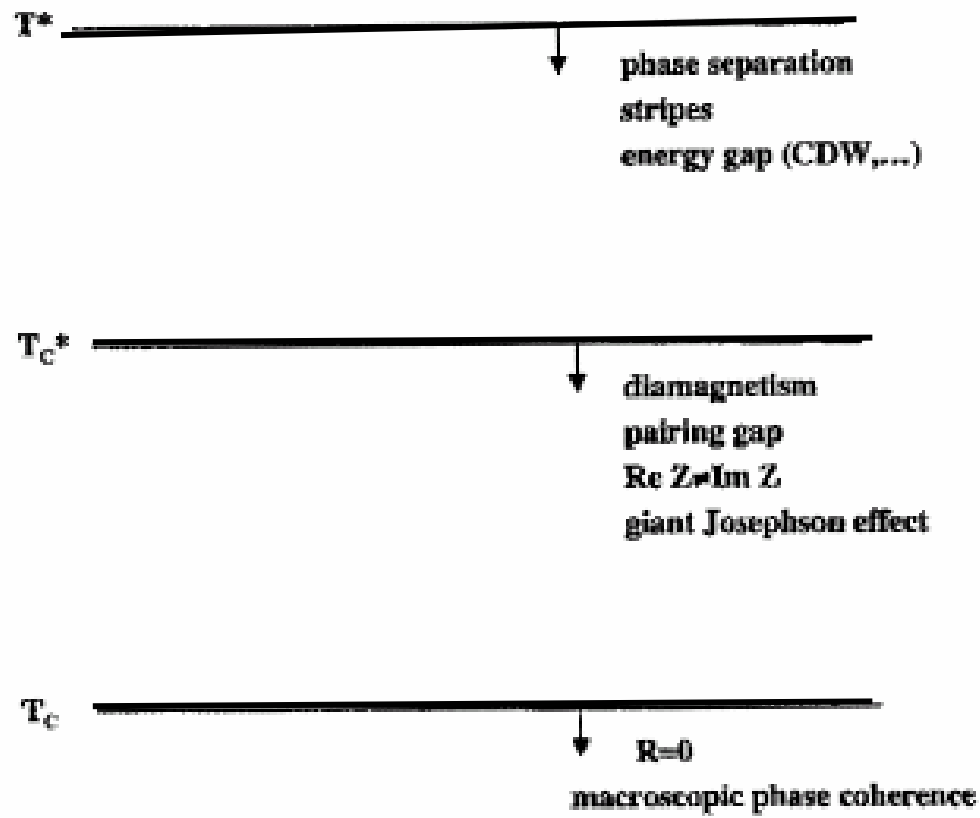

Figure 10.

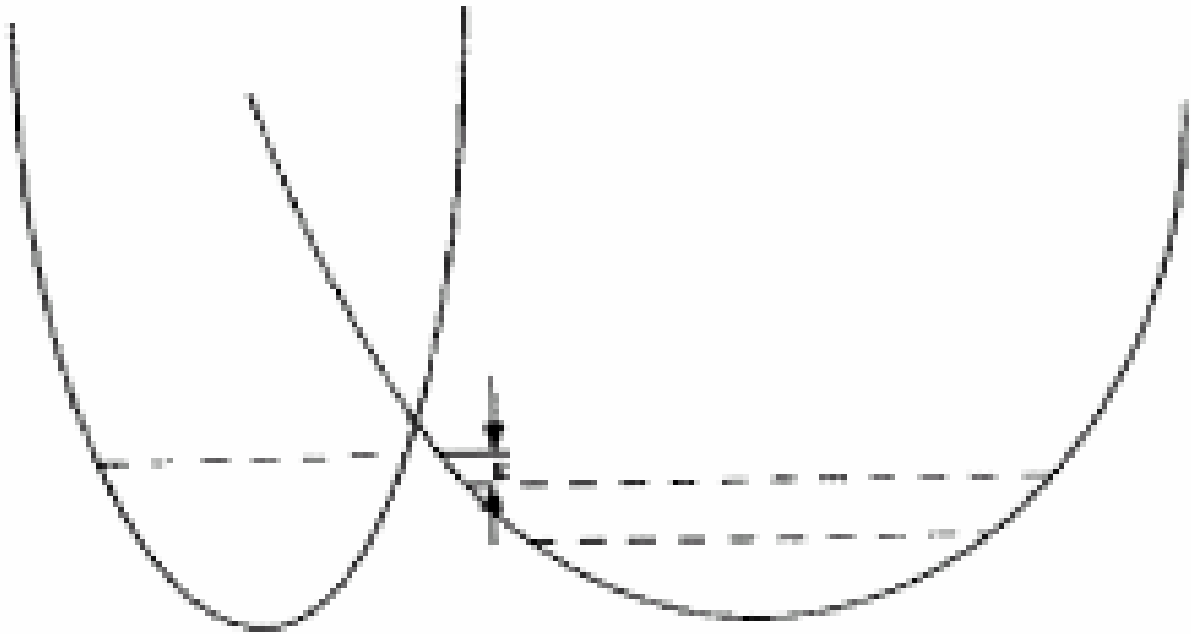

**Crossing of Terms**

Figure 11.

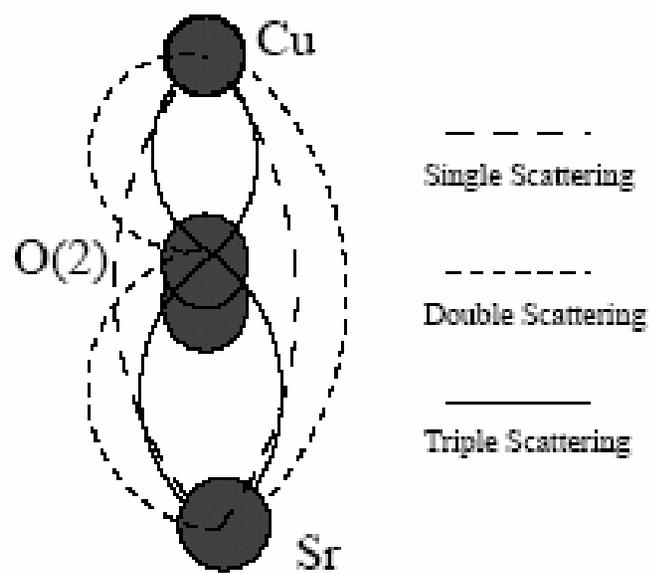

Figure 12.

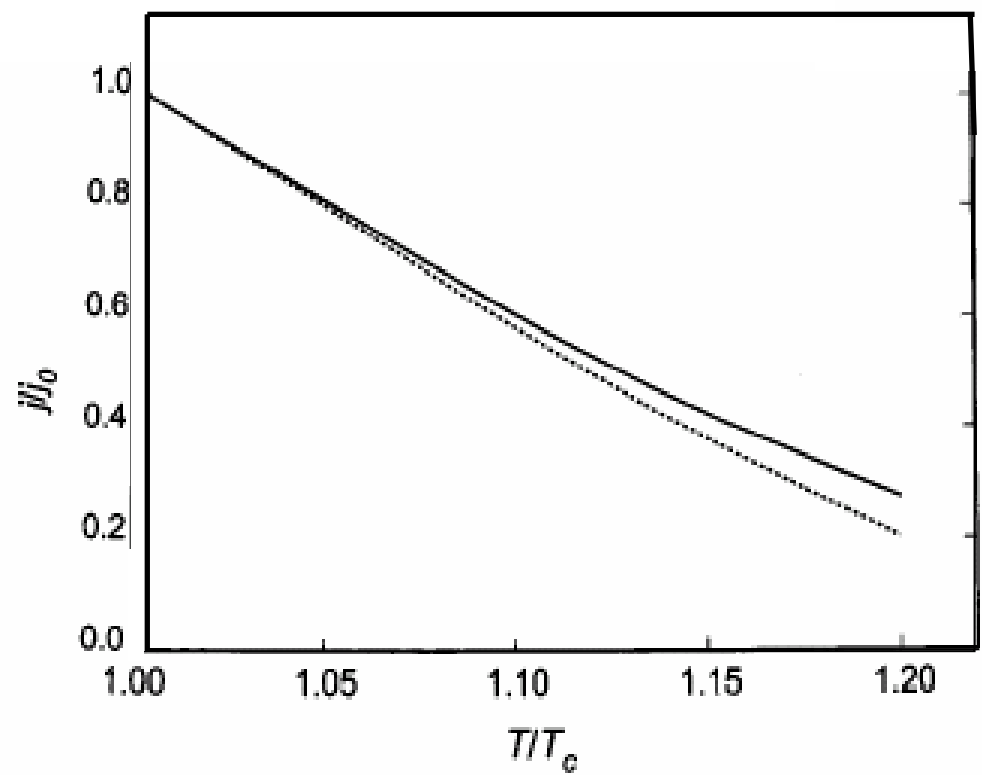

Figure 13.

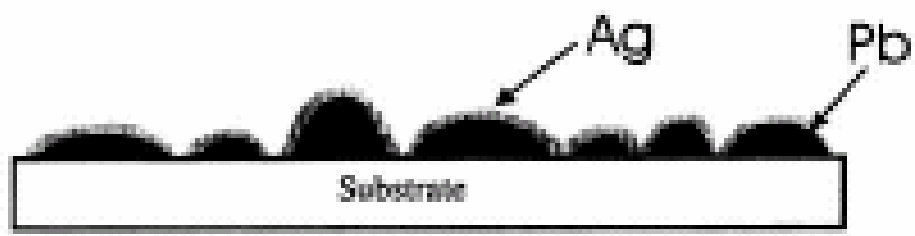